\documentclass[showpacs,aps,twocolumn,amssymb,pra]{revtex4}

\usepackage{bm}
\usepackage{epsfig}
\usepackage{amsmath} 
\usepackage{float}
\usepackage{verbatim}

\def\e{\epsilon}

\def\calM{\mathcal M}

\def\beq{\begin{equation}}
\def\eeq{\end{equation}}

\begin{document}

\title {Mean-field yrast spectrum of a two-component Bose gas in ring geometry: persistent currents at higher angular momentum}
\author{Z. Wu and E. Zaremba}
\affiliation{Department of Physics, Queen's University,
Kingston, ON,Canada}
\date{\today}
\begin{abstract}
We use analytic soliton solutions of a two-component Bose gas in the ring geometry to analyze the mean-field yrast spectrum of the system. We find that the spectrum exhibits a surprisingly rich structure as a result of an intricate interplay of interparticle interactions and population imbalance. We discuss the implication of these results in regard to the possibility of persistent currents at higher angular momenta. 
\end{abstract}
\pacs{67.85.De, 03.75.Kk, 03.75.Mn, 05.30.Jp}
\maketitle

\section{Introduction}
According to Bloch~\cite{Bloch}, the possibility of persistent currents in a
bosonic system in the ring geometry is determined by the {\it
yrast} spectrum~\cite{Mottelson}.
which is defined by the lowest energy state for
a given total angular momentum $L$. For a single-species system,
Bloch~\cite{Bloch} showed that the yrast spectrum takes the form
\beq
E_0(L) = \frac{L^2}{2M_TR^2} + e_0(L),
\eeq
where $M_T = NM$ is the total mass of the system, $R$ is the
radius of the ring on which the bosons reside, and $e_0(L)$ is a
periodic function of the angular momentum with the properties
$e_0(-L) = e_0(L)$ and $e_0(L+\nu N \hbar) = e_0(L)$. Here, $N$
is the number of particles of mass $M$ and $\nu$ is an integer. Bloch argued that the
system can support persistent currents at some non-zero angular
momentum $L_0$ if $E_0(L_0)$ is a local minimum.

Bloch's analysis can be extended to a binary mixture containing
$N_A$ particles of mass $M_A$ and $N_B$ particles of mass $M_B$.
In general, the function $e_0(L)$, where $L$ is again the total
angular momentum, exhibits a periodicity only if the mass ratio
$M_A/M_B$ is a rational number. For the special case $M_A =
M_B$, which is realizable with bosons in different hyperfine
states, $e_0(L)$ retains the same periodicity as for the
single-species system, but with $N=N_A+N_B$. In terms of $l=L/N\hbar$, the angular momentum per particle in
units of $\hbar$, one has $e_0(l+\nu) = e_0(l)$ and $e_0(-l)=e_0(l)$.
In this case, the
behaviour of $e_0(l)$ in the interval $0\le l\le 1/2$ is
sufficient to decide on the possibility of persistent currents.

For a single-spieces system of bosons interacting via a contact
interaction, the yrast spectrum was deterimined by Lieb~\cite{Lieb}. The
spectrum is defined by the so-called type-II excitations which
have been identified as soliton-like by comparing
their energy with those of soliton states as obtained
from a mean-field analysis~\cite{Ishikawa,Kanamoto1,Kanamoto2}. Although a similar identification
has not been established for the binary system, it has generally
been assumed that a mean-field analysis would provide a
reasonably accurate approximation of the yrast spectrum in this 
case as well. 

Determination of such a mean-field spectrum requires the knowledge
of soliton solutions to coupled Gross-Pitaevskii equations.
General solutions of this kind, expressed in terms of 
Jacobi elliptic functions,
were first obtained by Porubov and Parker~\cite{Porubov}
in the context of solving a Manokov system of coupled nonlinear 
Schr{\"o}dinger equations. Smyrnakis {\it et al.}~\cite{Smyrnakis3,Smyrnakis2} 
found these same solutions in the context
of the binary bosonic mixture in the ring geometry and applied 
them to determine the mean-field yrast spectrum of a system
with relatively weak interaction strength and small population 
imbalance. One of the purposes of this paper is to present a complete
analysis of the mean-field yrast spectrum of a binary 
system with {\it arbitrary} interaction strength and population
imbalance. We find that the yrast spectrum reveals a surprisingly
rich structure, not seen in Refs.~\cite{Smyrnakis3,Smyrnakis2}, as a 
result of the interplay of interaction and population imbalance.
%In particular, we demonstrate the emergence of distinct soliton-like 
%states in different ranges of the angular momentum as one
%decreases the concentration of the minority component 
%and increases the interaction strength. 

More importantly, our findings of the yrast spectrum provide
a definitive answer to the question as to whether persistent 
currents occur at higher angular momenta. The first study 
 of this kind for a binary system was carried out in Ref.~\cite{Smyrnakis1} 
 wherein the authors
determined a limited portion of the yrast spectrum. With this
information, the authors concluded that an arbitrarily small
concentration of the minority component, $x_B$, would disrupt the
possibility of persistent currents at higher angular momenta. On
the other hand, we argued in a recent paper~\cite{Anoshkin} that the yrast 
spectrum must revert, on the basis of continuity, to that of 
the single-species system in the $x_B \to 0$ limit. A
semi-quantitative calculation of the yrast spectrum supported
this physically obvious expectation, but a definitive explanation
of how this occurred was not provided.

The second purpose of the present paper is to analyze the $x_B \to 0$
limit in more detail using the analytic soliton solutions.
This analysis confirms that the single-species results are
recovered in the $x_B \to 0$ limit. However, this limit is
reached in a surprisingly complex way with the emergence of
distinct soliton-like states in different ranges of the angular
momentum. Furthermore, we obtain a simple, analytic formula
for the critical interaction strength above which persistent currents
are shown to exist at arbitrarily high angular momenta. This 
provides an explicit affirmation of the general argument 
we gave in Ref.~\cite{Anoshkin}.

The paper is organized as follows. In Sec. II we review the
soliton solutions of the GP equations for a binary mixture. In
Sec. III we present a detailed analysis of the mean-field yrast spectrum
for arbitrary interaction strengths and minority concentrations $x_B$. This information
is then used to investigate the possibility of persistent
currents at higher angular momenta and explain in detail how the
single-species limit is recovered for $x_B \to 0$. Our
conclusions are presented in the final section.\section{Two-component model and the ansatz soliton-train solutions}
We  consider an equal-mass, two-component Bose gas in the
one-dimensional 
ring geometry with repulsive interactions described by the following 
Hamiltonian
\begin{equation}
\hat H=-\sum_{is}\frac{\hbar^2}{2M R^2}\frac{\partial^2 }{\partial 
\theta_{si}^2} +\sum_{ss'}\sum_{ij} \frac{U}{4\pi R}\delta 
(\theta_{si}-\theta_{s'j}),
\end{equation}
where $U$ is the effective interaction strength and $s=A,B$ 
distinguishes the two atomic species.
The numbers of atoms in the two components, $N_A$ and $N_B$ 
respectively, are fixed and for definiteness we take $N_A \geq 
N_B$. 

Within mean-field theory, the system is described in terms of 
condensate wave functions $\psi_A(\theta)$ and $\psi_B(\theta)$ which 
are taken to have unit 
normalization $\int_0^{2\pi}d\theta |\psi_s(\theta)|^2 =1$. The 
mean-field energy functional, in units of $N\hbar^2/(2MR^2)$ where 
$N=N_A+N_B$, is given 
by  
\begin{align}
\bar E[\psi_A,\psi_B]=\sum_{s}x_s\int_0^{2\pi}d\theta \left 
|\frac{d\psi_s}{d\theta}\right |^2 +\pi\gamma \int_0^{2\pi}d\theta 
\rho^2(\theta),
\label{en_fun}
\end{align} 
where $x_s=N_s/N$, $\gamma=NMR^2U/\pi\hbar^2$ and 
$\rho(\theta)=x_A|\psi_A|^2+x_B|\psi_B|^2$ is the normalized
number density. Our objective is to
minimize the energy with respect to $\psi_s(\theta)$ subject to
the two normalization contraints and the angular momentum constraint
\begin{equation}
\bar L \equiv \frac{1}{i}\sum_{s}x_s\int_0^{2\pi}d\theta 
\psi_s^*\frac{d}{d\theta}\psi_s =l.
\label{agmcon}
\end{equation}
This can be achieved by minimizing the functional
\begin{equation}
\bar F[\psi_A,\psi_B] = \bar E - \Omega \bar L -\sum_s x_s
\mu_s \int_0^{2\pi} d\theta |\psi_s(\theta)|^2,
\label{f_fun}
\end{equation}
where $\Omega$ and $\mu_s$ are (dimensionless) 
Lagrange multipliers. The
variation of $\bar F$ with respect to $\psi_s$ leads to the
coupled time-independent Gross-Pitaevskii (GP) equations
\begin{equation}
-\psi''_s(\theta)+i \Omega\psi'_s(\theta)+2\pi\gamma \rho 
(\theta)\psi_s(\theta)= \mu_s \psi_s(\theta).
\label{gpe}
\end{equation}
These same equations can be obtained from the {\it time-dependent}
GP equations by assuming solutions of the form $\psi_s(\theta,t)
=\psi_s(\theta-\Omega t)e^{-i\mu_s t}$, where
$\Omega$ is the angular 
velocity of the soliton, in units of $N\hbar/(2MR^2)$ . 
Although Eq.~(\ref{gpe}) can be viewed as 
representing the system in a rotating frame of reference in 
which the soliton is at rest, we emphasize that $\Omega$ is
{\it not} a free parameter but is determined by the soliton
state itself.

The lowest energy soliton solution consistent with the angular
momentum constraint in Eq.~(\ref{agmcon}) defines the mean-field
yrast spectrum~\cite{Mottelson}, $\bar E_0(l)$, as a function of
the angular momentum per particle, $l$. The fact that the
soliton is a variational minimum of Eq.~(\ref{f_fun}) leads to 
the relation
\begin{equation}
\Omega = \frac{\partial \bar E_0(l)}{\partial l}.
\label{omega}
\end{equation}
Eq.~(\ref{omega}) is an important and useful relation
which shows that the slope of the yrast spectrum $\bar E_0 (l)$ 
is simply the angular velocity $\Omega$.

To solve the coupled equations in Eq.~(\ref{gpe}) 
%can be solved via an 
%ansatz~\cite{Porubov, Smyrnakis2}. For completeness and also for the 
%purpose of the following 
%development, we will provide here a derivation of these solutions. To 
%do so, one makes use of the parameterization 
we make use of the modulus-phase representation
\begin{equation}
\psi_s(\theta)=\sqrt{\rho_s(\theta)}e^{i\varphi_s(\theta)}.
\label{2para}
\end{equation} 
The single-valuedness of the wave function $\psi_s(\theta)$ implies the 
following boundary conditions 
\begin{align}
\label{2bdc1}
\rho_s(\theta+2\pi)-\rho_s(\theta)&=0 \\
\varphi_s(\theta+2\pi)-\varphi_s(\theta)&=2\pi J_s, \quad J_s=0,\pm 
1,\pm 2,\cdots
\label{2bdc2}
\end{align}
where the integers $J_s$ are referred to as phase winding numbers.

Substituting Eq.~(\ref{2para}) into Eq.~(\ref{gpe}), one obtains
\begin{equation}
-\left(\sqrt{\rho_s}\right)''+\left (\varphi'_s -\Omega\right 
)\varphi'_s\sqrt{\rho_s}+2\pi\gamma\rho \sqrt{\rho_s}=\mu_s
\sqrt{\rho_s}
\label{ampeqa}
\end{equation}
and
\begin{equation}
\sqrt{\rho_s}\varphi''_s+\left (2\varphi'_s-\Omega\right )\left( 
\sqrt{\rho_s}\right )'=0.
\label{pheqa}
\end{equation}
The solution to Eq.~(\ref{pheqa}) is given by 
\begin{equation}
\varphi'_s(\theta)=\frac{W_s}{2\rho_s}+\frac{\Omega}{2},
\label{pheq_sola}
\end{equation}
where $W_s$ is an integration constant to be determined. 
Physically, $\varphi'_s(\theta)$ gives the superfluid velocity field of 
each component.
%and should not be confused with the angular 
%velocity $\Omega$ with which the 
%soliton travels. 
Using Eq.~(\ref{agmcon}) and Eq.~(\ref{pheq_sola}), one 
finds that the angular momentum per particle can be written as 
\begin{equation}
l=x_Al_A+x_Bl_B,
\label{agconsab}
\end{equation}
where 
\begin{equation}
l_s=\pi W_s+\frac{\Omega}{2}
\label{agms}
\end{equation}
is the angular momentum per particle of each species. Finally, 
inserting Eq.~(\ref{pheq_sola}) into Eq.~(\ref{ampeqa}), one obtains
\begin{equation}
\frac{1}{2}\rho_s\rho''_s-\frac{1}{4}\left(\rho'_s\right)^2 
-2\pi\gamma\rho\rho_s^2 +\tilde\mu_s \rho_s^2-\frac{W_s^2}{4}=0,
\label{ampeq2a}
\end{equation}
where $\tilde \mu_s=\mu_s+{\Omega}^2/4$. These are the coupled 
equations that determine the densities of the two components. 

%Let us try the following ansatz~\cite{Smyrnakis1} for 
The coupled equations in Eq.~(\ref{ampeq2a}) admit solutions of
the form~\cite{Porubov, Smyrnakis3,Smyrnakis2}
\begin{equation}
\rho_B=c_B\left (1+c_A^{-1}\rho_A\right ),
\label{ansatz}
\end{equation}
where $c_s$ are coefficients to be determined.
Since $\rho_A$ and $\rho_B$ are normalized to unity, an angular
integration of Eq.~(\ref{ansatz}) yields the following 
relationship between $c_A$ and $c_B$, 
\begin{equation}
2\pi c_B+c_Bc_A^{-1}=1.
\label{cacb1}
\end{equation}
Using the ansatz in Eq.~(\ref{ansatz}) to eliminate $\rho_B$ for $s=A$ 
and $\rho_A$ for $s=B$, respectively, from Eq.~(\ref{ampeq2a}), one 
finds
\begin{align}
\frac{1}{2}\rho_s\rho''_s-\frac{1}{4}\left(\rho'_s\right)^2 
-2\pi\gamma_s\rho_s^3 +\bar \mu_s\rho_s^2-\frac{W_s^2}{4}=0.
\label{rhoseq}
\end{align}
Here 
\begin{align}
\gamma_s&=c_s^{-1}\left(c_Ax_A+c_Bx_B\right ) \gamma,
\label{gammas}
\end{align}
and
\begin{align}
\bar \mu_s=\tilde \mu_s \mp 2\pi\gamma  c_{\bar s}x_{\bar s},
\end{align}
where $\bar s$ denotes the complementary species to $s$ and 
the $-$ ($+$) sign  
corresponds to the $A$ ($B$) component. It is clear 
from Eq.~(\ref{gammas}) 
that the effective interactions $\gamma_s$ depend on the
coefficients $c_s$ only only through the ratio
$r\equiv c_B/c_A$ and that $\gamma_A/\gamma_B = r$. Having
introduced this ratio, we observe that the normalization
relation in Eq.~(\ref{cacb1}) implies that
\beq
c_A = \frac{1-r}{2\pi r},\quad c_B = \frac{1-r}{2\pi},
\label{coeff_r}
\eeq
and that the density ansatz in Eq.~(\ref{ansatz}) takes the form
\beq
\rho_B = \frac{1-r}{2\pi} + r \rho_A.
\label{ansatz_r}
\eeq
The ratio $r$ plays a central role in the subsequent analysis.

Equation (\ref{rhoseq}) 
indicates that the ansatz in Eq.~(\ref{ansatz}) reduces the 
coupled system to two {\it independent} equations, one for each
species with a modified effective interaction. Although
independent, the solutions must nevertheless be consistent with
the starting ansatz. As we shall see, there is indeed sufficient
freedom in the form of the two independent solutions to ensure
that Eq.~(\ref{ansatz}) is satisfied. We refer to this condition
as the self-consistency requirement.
(For an alternative method of checking self-consistency, see 
Refs.~\cite{Porubov, Smyrnakis3,Smyrnakis2}.) 
We finally note that the 
ansatz allows one to express the energy per particle as
\begin{equation}
\bar E =x_A\bar E_A +x_B\bar E_B-2\pi^2 \gamma x_Ax_Bc_Ac_B,
\label{E_total}
\end{equation}
where 
\begin{equation}
\bar E_s =\int_0^{2\pi}d\theta \left |\frac{d\psi_s}{d\theta}\right |^2 
+\pi\gamma_s \int_0^{2\pi}d\theta \rho_s^2(\theta).
\end{equation}
Despite its appearance, Eq.~(\ref{E_total}) accounts fully for the
interactions between the two atomic species.

There are two classes of solutions to Eq.~(\ref{rhoseq}), one 
corresponding to $ c_Ax_A+c_Bx_B=0$ and the other to
$ c_Ax_A+c_Bx_B\neq 0$. We analyze these two classes in
turn.
\subsubsection{Class (i) solutions: $ c_Ax_A+c_Bx_B=0$}
In this case, $\gamma_A=\gamma_B=0$ and  the system is 
effectively 
reduced to two {\it non-interacting} species. Since $r =
-x_A/x_B$, Eq.~(\ref{coeff_r}) gives
\beq
c_A=- \frac{1}{2\pi x_A},\quad c_B= \frac{1}{2\pi x_B}
\eeq
and Eq.~(\ref{ansatz_r}) implies that the total density is a
constant, $\rho(\theta) = 1/2\pi$. For $\gamma_s =0$,
Eq.~(\ref{rhoseq}) becomes
 \begin{align}
\label{scasp}
\frac{1}{2}\rho_s\rho''_s-\frac{1}{4}\left(\rho'_s\right)^2  +\bar\mu_s 
\rho_s^2-\frac{W_s^2}{4}=0.
\end{align}
Defining the function $y_s = -\rho_s''/4+\bar \mu_s \rho_s^2
-W_s^2/4$, Eq.~(\ref{scasp}) is equivalent to
\beq
\frac{y_s'}{y_s} = \frac{\rho_s'}{\rho_s}
\eeq
which, via an integration, leads to
\begin{equation}
\frac{1}{4}\left(\rho'_s\right)^2+\bar\mu_s 
\rho_s^2+V_s\rho_s+\frac{W_s^2}{4}=0,
\label{rhoeqni}
\end{equation}
where $V_s$ is an integration constant.

Using the boundary condition in Eq.~(\ref{2bdc1}) and 
the normalization of $\rho_s$, we find that the solution to 
Eq.~(\ref{rhoeqni}) is given 
by
\begin{equation}
\rho_s(\theta)=\frac{1}{2\pi}\left [1+d_s\cos
j_s(\theta-\theta_{0s})\right ],
\label{rhosolni}
\end{equation}
where $j_s=1,2,\cdots$. Substituting this solution back into 
Eq.~(\ref{rhoeqni}), we find
\begin{align}
\bar \mu_s=\frac{j_s^2}{4}, \qquad V_s&=-\frac{j_s^2}{4\pi}
\end{align}
and
\begin{align}
W_s^2&=\frac{j_s^2}{4\pi^2}\left( 1- d_s^2 \right ). 
\label{W_sni}
\end{align}
The undetermined constants $d_s$, $j_s$ and $\theta_{0s}$ are finally
determined by inserting Eq~(\ref{rhosolni}) into
Eq.~(\ref{ansatz_r}). For consistency, we require
$\theta_{0A}=\theta_{0B}=\theta_0$, $j_A = j_B =j$ and 
\begin{equation}
\frac{d_B}{d_A} = r =-\frac{x_A}{x_B}.
\label{dadb}
\end{equation}
In view of the arbitrary phase angle $\theta_0$, we can without loss of generality
choose $d_B$ to be positive.

Once the densities are known, the phases can be
determined from Eq.~(\ref{pheq_sola}). An angular integration of
this equation 
leads to the general phase-density relation
\begin{align}
\varphi_s(\theta)-\varphi_s(\theta_0)=\frac{\Omega}{2}(\theta 
-\theta_0)+\frac{W_s}{2}\int_{\theta_0}^\theta 
\frac{d\theta'}{\rho_s(\theta')}.
\label{phase_int}
\end{align}
Using Eq.~(\ref{rhosolni}) and writing
$j(\theta-\theta_0)=n\pi+\bar\theta$, where $n$ 
is a non-negative integer and $0\leq\bar\theta < \pi$, one finds
\begin{align}
&\varphi_s(\theta)-\varphi_s(\theta_0)-\frac{\Omega}{2}(\theta 
-\theta_0) \nonumber \\
=&\frac{{\rm sgn}W_s}{2}\left \{ {n\pi}+2\tan^{-1}\left [ 
\left(\frac{1+d_s}{1-d_s}\right 
)^{\pm\frac{1}{2}}\tan\frac{\bar\theta}{2}\right ] \right \},
\label{phases_ni}
\end{align}
where the $+$($-$) sign corresponds to even (odd) $n$ values. 
Finally, to evaluate the phase boundary condition in
Eq.~(\ref{2bdc2}), we insert $\theta - \theta_0 =2\pi$ (which
implies $n=2j$ and $\bar \theta =0$) in Eq.~(\ref{phases_ni}).
This gives immediately 
\begin{align}
{\Omega}&=2J_s-j\,{{\rm sgn}W_s},
\label{phase_bdni}
\end{align} 
which determines the value of $\Omega$. We see that the angular 
velocity for these solutions is a constant, independent of the 
parameter $d_s$ which 
specifies a particular solution. Eliminating $\Omega$ from 
Eq.~(\ref{phase_bdni}) we find
\begin{equation}
J_B-J_A=\frac{j}{2}\left ({{\rm sgn}W_B}-{{\rm sgn}W_A} \right )
\label{phasebcni}
\end{equation}
for the solutions given by Eq.~(\ref{rhosolni}) and 
Eq.~(\ref{phases_ni}).
\subsubsection{ Class (ii) solutions: $c_Ax_A+c_Bx_B\neq  0$}
In this case Eq.~(\ref{rhoseq}) 
admits the following soliton train solutions~\cite{Carr1,Carr2,Kanamoto1}
\begin{align}
\label{rhoa}
\rho_s(\theta)=\mathcal N_s(\eta_s)\left[1+\eta_s{\rm dn}^2\left(\left 
.\frac{j_sK}{\pi}(\theta-\theta_{0s})\right |m_s\right)\right ],
\end{align}
where ${\rm dn}(u|m)$ is a Jacobi elliptic function~\cite{Abramowitz} and $j_s= 1,2,\cdots$.
Here  
\begin{align}
\eta_s=\frac{-2j_s^2K^2(m_s)}{g_s}, \qquad \mathcal 
N_s(\eta_s)=\frac{g_s}{2\pi^3\gamma_s},
\label{eta_1}
\end{align}
and 
\begin{equation}
g_s\equiv \pi^2\gamma_s +2j_s^2 K(m_s)E(m_s),
\end{equation}
where $K(m)$ and $E(m)$ are, respectively, complete elliptic 
integrals of the first and 
second kind with elliptic parameter $m$~\cite{Abramowitz}. We point out that this form of 
solution 
applies to both bright (attractive interactions) and grey/dark 
(repulsive interactions) solitons. In addition, these
single-species solitons are completely specified by the elliptic
parameter $m_s$, the interaction parameter $\gamma_s$ and the
soliton train index $j_s$.
Substituting the solutions in 
Eq.~(\ref{rhoa}) back into Eq.~(\ref
{rhoseq}), one finds~\cite{Kanamoto1}
\begin{align}
\bar\mu_s=\frac{3g_s}{2\pi^2}-\frac{(2-m_s)j_s^2K^2(m_s)}{\pi^2}.
\end{align}
and
\begin{align}
{W_s^2}=\frac{g_sf_sh_s}{2\pi^8\gamma_s^2},
\label{Wsqr}
\end{align}
where $ f_s\equiv g_s-2j_s^2K^2(m_s)$ and $h_s\equiv
g_s-2(1-m_s)j_s^2K^2(m_s)$. 

We turn next to the implications of the density ansatz in
Eq.~(\ref{ansatz_r}).
Substituting Eq.~(\ref{rhoa}) into
Eq.~(\ref{ansatz_r}), we observe first that consistency requires 
$\theta_{0A}=\theta_{0B}=\theta_0$, $m_A = m_B = m$ and $j_A=j_B
=j$. More remarkably, one can check that Eq.~(\ref{ansatz_r}) is
satisfied identically for any value of the parameter $r$
provided the interaction parameters are related by
Eq.~(\ref{gammas}). Thus the pair of soliton train
solutions in Eq.~(\ref{rhoa}) are indeed solutions to 
Eq.~(\ref{ampeq2a}). To specify these solutions completely,
however, we must still determine the parameters $r$ and $m$.

To do so we must next make use of the phase boundary conditions
in Eq.~(\ref{2bdc2}). Inserting the densities in 
Eq.~(\ref{rhoa}) into Eq.~(\ref{phase_int}), one finds that the phases 
are given by
\begin{align}
\varphi_s(\theta)-\varphi_s(\theta_0)=\frac{\Omega}{2}(\theta 
-\theta_0)+\frac{\pi^4\gamma W_s}{ jf_sK}\Pi\left(n_s;u|m\right ),
\label{phase}
\end{align}
where $n_s=-2j^2mK^2/f_s$, $u=jK(\theta-\theta_0)/\pi$ and 
$\Pi\left(n_s; u|m\right )$ is the elliptic integral of the third kind. 
The phase boundary conditions then yield
\begin{equation}
{\Omega}=2J_s-2\mathcal M_s\, {\rm sgn} W_s,
\label{phasebc}
\end{equation}
where $\mathcal M_s =\sqrt{2g_sh_s/{f_s}}\Pi\left(n_s;K|m\right )/(2\pi 
K)$. 
Eliminating $\Omega$ from Eq.~(\ref{phasebc}) we obtain
\begin{equation}
J_B-J_A=\mathcal M_B{\rm sgn} W_B-\mathcal M_A{\rm sgn} W_A.
\label{phasebc'}
\end{equation}
This is the key equation that determines $r=c_B/c_A$ as a function of 
the elliptic parameter $m$ for this class of solutions. In other
words, once the relationship between $r$ and $m$ is established,
the solutions of physical interest are generated by allowing $m$ 
to vary over some well-defined interval to be determined.

\section{Mean-field yrast spectrum}
In this section, we use the phase boundary conditions, 
Eq.~(\ref{phasebcni}) and Eq.~(\ref{phasebc'}), to determine the 
mean-field yrast spectrum $\bar 
E_0(l)$.  As shown in~\cite{Smyrnakis1, Anoshkin} it is
advantageous to consider the energy $\bar e_0 (l)\equiv \bar E_0 -l^2$, 
which is a periodic function of the angular 
momentum per particle $l$, with unit period. In addition, 
$\bar e_0 (l)$ has the inversion property $\bar e_0(-l)=\bar e_0(l)$, 
which reflects 
the fact that the energy does not depend on the sense of the
angular momentum. It is thus sufficient to restrict our discussion of 
the yrast 
spectrum to $0\leq l \leq 1/2$. Finally, as the soliton solutions with 
$j > 1$ generally have higher energies, we will focus on the case of $j 
=1$, namely the 
single-soliton states. 

Before we consider the two classes of solutions individually, it is 
useful to observe that both phase boundary conditions, 
Eqs.~(\ref{phasebcni}) and (\ref{phasebc'}), involve only the difference 
of the phase winding numbers $\mathcal J \equiv J_B - J_A$. From  
Eq.~(\ref{phase_bdni}) we see that a simultaneous shift
of both $J_A$ and $J_B$ by an integer $p$ simply changes 
the angular velocity from $\Omega$ to $\Omega + 2p$ and,  in view of 
Eqs.~(\ref{agconsab}) and (\ref{agms}),  the angular momentum from 
$l$ to $l + p$. In addition, a simultaneous sign reversal of
$J_A$ and $J_B$, along 
with that of $W_A$ and $W_B$, changes the angular velocity from 
$\Omega$ to $-\Omega$ and the angular momentum from $l$ to $-l$.

We now consider the class of solutions obtained for $c_Ax_A+c_Bx_B = 
0$, given by Eq.~(\ref{rhosolni}) and Eq.~(\ref{phases_ni}). These 
solutions will 
be shown to represent those states for which the angular momentum per 
particle lies in the range $0\leq l \leq x_B $ (in the interval 
$0\leq l \leq 1/2$ of interest). To 
begin, we observe from the boundary condition in Eq.~(\ref{phasebcni}), 
that $\mathcal J = J_B -  J_A$ can only take the value $0$ and $\pm 1$ 
for $j=1$. We find 
that the following two combinations of $J_A$ and $J_B$ yield angular 
momenta that lie within $0\leq l \leq 1/2$:  (i) $J_A=J_B=0$ with ${\rm 
sgn W_A}=
{\rm sgn W_B}=-1$; (ii) $J_A=0$ and $J_B=1$ with ${\rm sgn}W_A=-1$ and 
${\rm sgn}W_B=1$.  From Eq.~(\ref{phase_bdni}) we find that $\Omega =1$ 
for 
both (i) and (ii). 

A combination of Eqs.~(\ref{agms}), (\ref{W_sni}) and
(\ref{phase_bdni}) shows that the angular momentum of each
species can be expressed as
\beq
l_s =J_s + \frac{1}{2}{\rm sgn}W_s (\sqrt{1-d_s^2}-1).
\label{ang_mom}
\eeq
For $J_s=0$ and $d_s = 0$ we have a uniform density state of
zero angular momentum. In order for $l_s$ to increase to
positive values as $|d_s|$ increases from zero, we must
choose ${\rm sgn}W_s$ to be negative. This establishes parameter set
(i). Since $|d_B| > |d_A|$, the allowable range of $|d_B|$
values is set by the requirement that $W_B^2$ is non-negative.
This determines that $0\le |d_B| \le 1$. As $|d_B|$ increases
continuously from 0 to 1, Eq.~(\ref{ang_mom}) shows that $l_A$
increases from 0 to $(x_A-\sqrt{x_A-x_B})/2x_A$ and $l_B$
increases from 0 to 1/2. The total angular momentum per particle as a 
result increases from zero to $\left(1-\sqrt{x_A-x_B}\right )/2$. At 
this point, the density 
of the $B$ component acquires a node at $\theta=\theta_0+\pi$ (recall that 
$d_B$ is chosen to be positive) and the phase 
winding number $J_B$ jumps by 1 as $l$ increases further. By the
same token ${\rm sgn}W_B$ must become positive in order that
$l_B$ be continuous with further variations of $|d_B|$. On the
other hand $J_A$ and ${\rm sgn} W_A$ retain their values of 0
and $-1$, respectively, thus defining parameter set (ii).
As $|d_B|$ is reduced continuously back to zero, 
$l_A$ decreases from $(x_A-\sqrt{x_A-x_B})/2x_A$ to zero 
while $l_B$ 
continues to increases from $1/2$ to $1$, at which point all the 
circulation is carried by the $B$ component and the total
angular momentum is $l = x_B$. Using 
Eq.~(\ref{phases_ni}), we evaluate 
the phases of the components at this point and find $\varphi_A(\theta) 
-\varphi_A(\theta_0) 
= 0$  and $\varphi_B(\theta) - \varphi_B(\theta_0) 
= \theta-\theta_0$.  This shows that the components are in
the plane wave states $\psi_A = 1/\sqrt{2\pi}$ and $\psi_B = 
e^{i(\theta-\theta_0)}/\sqrt {2\pi}$ when $l = x_B$.

To obtain the energy spectrum for the states given by 
Eq.~(\ref{rhosolni}) and Eq.~(\ref{phases_ni}), we can use the formula 
given in Eq.~(\ref{E_total}). 
However, a simpler way in this case is to make use of the fact that 
$\Omega = 1$ and integrate Eq.~(\ref{omega}). Recalling that the ground 
state energy in 
the mean-field theory is $\gamma/2$, we find
\begin{align}
\bar E_0=\frac{\gamma}{2}+l
\label{E0ni}
\end{align}
for $0\leq l \leq x_B$, which coincides with the yrast spectrum of an 
ideal gas. It can be shown~\cite{Smyrnakis1, Anoshkin} 
that the solutions in Eq.~(\ref{rhosolni}) and Eq.~(\ref{phases_ni}) in 
fact yield the 
lowest possible mean-field energy as a function of $l$. In other words, 
Eq.~(\ref{E0ni}) indeed gives the yrast spectrum for $0\leq l \leq x_B$.

The states with angular momenta within the range $x_B\leq l \leq 1/2$
belong to the second class of soliton solutions, given by 
Eq.~(\ref{rhoa}) and Eq.~(\ref{phase}). We again focus on the $j=1$ case and consider the 
boundary condition in Eq.~(\ref{phasebc'}), which determines $r$ as a 
function of the 
elliptic parameter $m$. Once $r(m)$ is determined, various physical 
quantities can then be evaluated using the formulae given in Sec. II. 
Unlike Eq.~(\ref
{phasebcni}) for the first class of solutions, however, 
Eq.~(\ref{phasebc'}) can only be solved numerically and generally 
admits solutions for $\mathcal J  > 
1$ even in the case of $j =1$. Furthermore, the solution $r= 
r(m,\mathcal J)$ is multi-branched for fixed values of $\mathcal J$, 
adding to the complexity of 
the problem. We will focus on branches with negative values of 
$r(m,\mathcal J)$ (grey-bright solitons), as these yield lower 
energies than the positive branches (grey-grey solitons). 

For orientation purposes, it is useful to first determine 
$r_0(\mathcal J)\equiv 
r(m=0,\mathcal J)$.  It is important to note that, should $m=0$ 
solutions exist, Eq.~(\ref{rhoa}) and Eq.~(\ref{phase}) imply
that the $A$ and $B$ components are both in plane wave states. 
More specifically, one obtains at 
these $r$ values 
the wave functions $\psi_A = e^{iJ_A (\theta-\theta_0)}/\sqrt{2\pi}$ 
and $\psi_B = e^{iJ_B (\theta-\theta_0)}/\sqrt{2\pi}$, for which the 
total angular 
momentum is $l = x_A J_A + x_B J_B$. Thus, to obtain solutions whose 
angular momenta are within $x_B\leq l \leq 1/2$,  $J_A$ clearly has to 
be zero 
while $J_B = \mathcal J$ can take on various positive integer 
values dependening on the size of $x_B$. 

Setting $m=0$ in Eq.~(\ref{phasebc'}), we find the simplified equation 
(see the Appendix for the calculation of $\mathcal M_s$)
\begin{equation}
\mathcal J=\frac{1}{2}{\rm sgn} W_B\sqrt{2\gamma_B +1}-\frac{1}{2}{\rm 
sgn} W_A\sqrt{2\gamma_A +1}.
\end{equation}
Using the expressions for $\gamma_s$ in Eq.~(\ref{gammas}), this
equation can be put into the form of the following quartic equation
for $r_0(\mathcal J)$
\begin{widetext}
\begin{align}
\gamma^2(x_Br_0 + x_A)^2 (r_0 - 1)^2 - 4 {\mathcal J}^2 \gamma r_0(r_0 
+ 1) (x_Br_0 + x_A) + 4 \mathcal J^2(\mathcal J^2 -1) r_0^2 = 0.
\label{r_0}
\end{align}
\end{widetext}
The number of negative solutions that this equation admits depends 
on the values of $\gamma$, $x_B$ and $\mathcal J$. In view of our
earlier discussion, the case of $\mathcal J = 0$ is not of interest 
when $l>x_B$. For $\mathcal J = 1$, it 
is not difficult to see that there are always two negative solutions, 
one of which is $r_0 = -
x_A/x_B$ for {\it arbitrary} $\gamma$ and $x_B$. This special
root for $\mathcal J = 1$ is associated with the class (i)
solutions considered earlier. The other corresponds to a
different soliton branch that begins when $l$ exceeds $x_B$.
 \begin{figure}[!htbp]
\begin{center} 
 \includegraphics[angle=0, width=1\columnwidth]{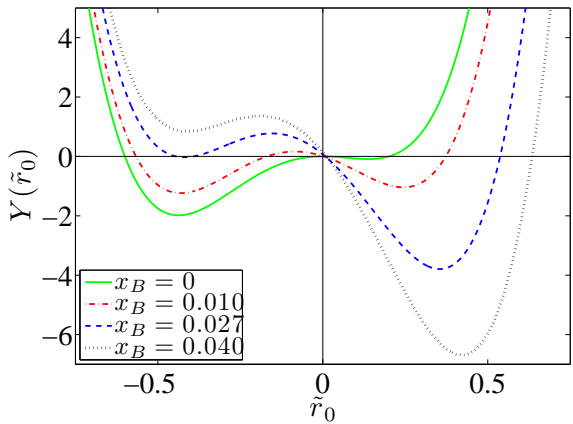}
 \caption
 {The quartic function $Y(\tilde r_0)$ plotted for various $x_B$ values 
for $\gamma=10$ and $\mathcal J = 2$.}
\label{Y_r}
\end{center}
\end{figure}
 \begin{figure}[!htbp]
\begin{center} 
 \includegraphics[angle=0, width=1\columnwidth]{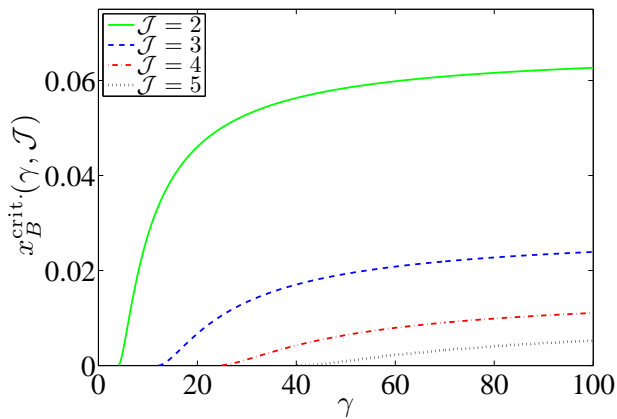}
 \caption
 {The critical value $x_B^{\rm crit}(\gamma,\mathcal J)$ plotted as a 
function of $\gamma$ for $\mathcal J=2$ (solid, green), 3
(dashed, blue), 4 (dot-dash, red) and 5 (dotted, black).}
\label{x_B_crit}
\end{center}
\end{figure}
The situation for 
$\mathcal J  > 1$ is slightly more complicated.  To analyze it, we 
multiply both sides of 
Eq.~(\ref{r_0})  by $x_B^2$ and rewrite it as an equation for $\tilde 
r_0 = x_B r_0$ 
 \begin{equation}
 Y(\tilde r_0) \equiv y(\tilde r_0) + z(\tilde r_0) = 0,
 \label{r_0II}
 \end{equation}
 where 
  \begin{equation}
 y(\tilde r_0) = \tilde r_0^2\left [\gamma \tilde r_0 + \gamma - 
2\mathcal J(\mathcal J -1) \right ]\left [\gamma \tilde r_0 + \gamma - 
2\mathcal J(\mathcal J +1) 
\right ]
 \end{equation}
 and
  \begin{align}
 z(\tilde r_0)& = -4x_B\gamma^2\tilde r_0^3  - 6 x_Ax_B \gamma^2 \tilde 
r_0^2 \nonumber \\
 &\quad+ 2x_A x_B \gamma [(x_B-x_A)\gamma -\mathcal J^2]\tilde r_0+ 
\gamma^2 x_A^2 x_B^2.
 \end{align}
The cubic $z(\tilde r_0)$ is proportional to $x_B$ and thus does
not influence the locations of the roots given by $y(\tilde r_0) =0$
in the $x_B \to 0$ limit. The 
quartic $y(\tilde r_0)$ has a double root at $\tilde r_0 =0$ 
and two single roots 
at $[-\gamma + 2\mathcal J(\mathcal J \pm1)]/\gamma$. It is clear that 
it has two negative roots if $\gamma > 2\mathcal J(\mathcal J +1)$, 
which guarantees that $Y(\tilde r_0) $ has
two negative roots for sufficiently 
small $x_B$. This condition, however, is actually too strong. If 
$2\mathcal J (\mathcal J -1) < \gamma < 2\mathcal J (\mathcal J
+1)$, $y(\tilde r_0)$ has one positive and one negative root
(Fig.~\ref{Y_r} gives a numerical example for $\mathcal J = 2$
and $\gamma = 10$). As $x_B$ is increased from 0, the cubic
$z(\tilde r_0) $ removes the root degeneracy at $\tilde r_0 = 0$ 
and two negative roots of $Y (\tilde r_0)$ appear. 
\begin{figure*}[!htbp]
\begin{center} 
 \includegraphics[angle=0, width=0.64\columnwidth]{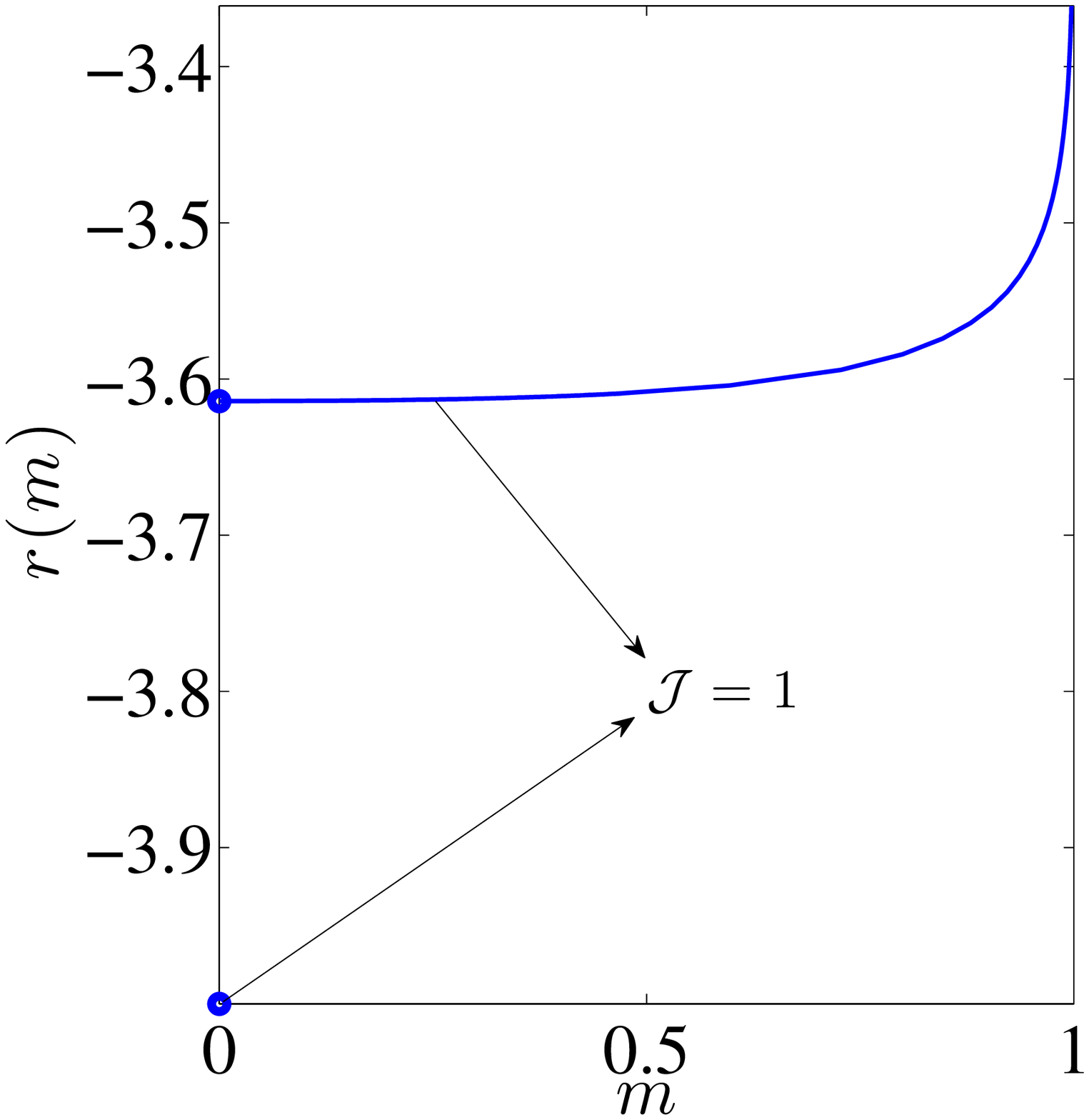}
  \includegraphics[angle=0, width=0.64\columnwidth]{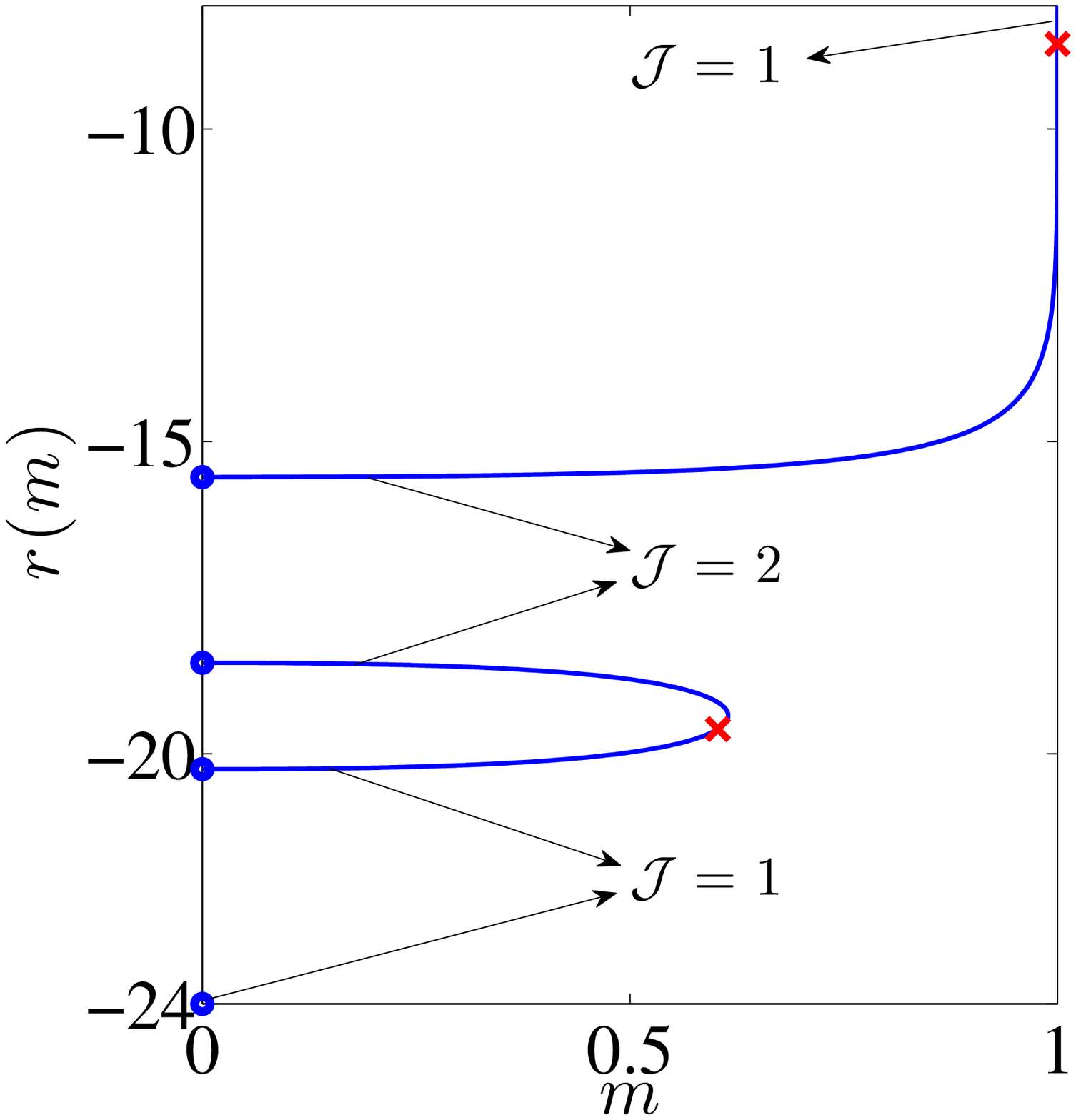}
 \includegraphics[angle=0, width=0.64\columnwidth]{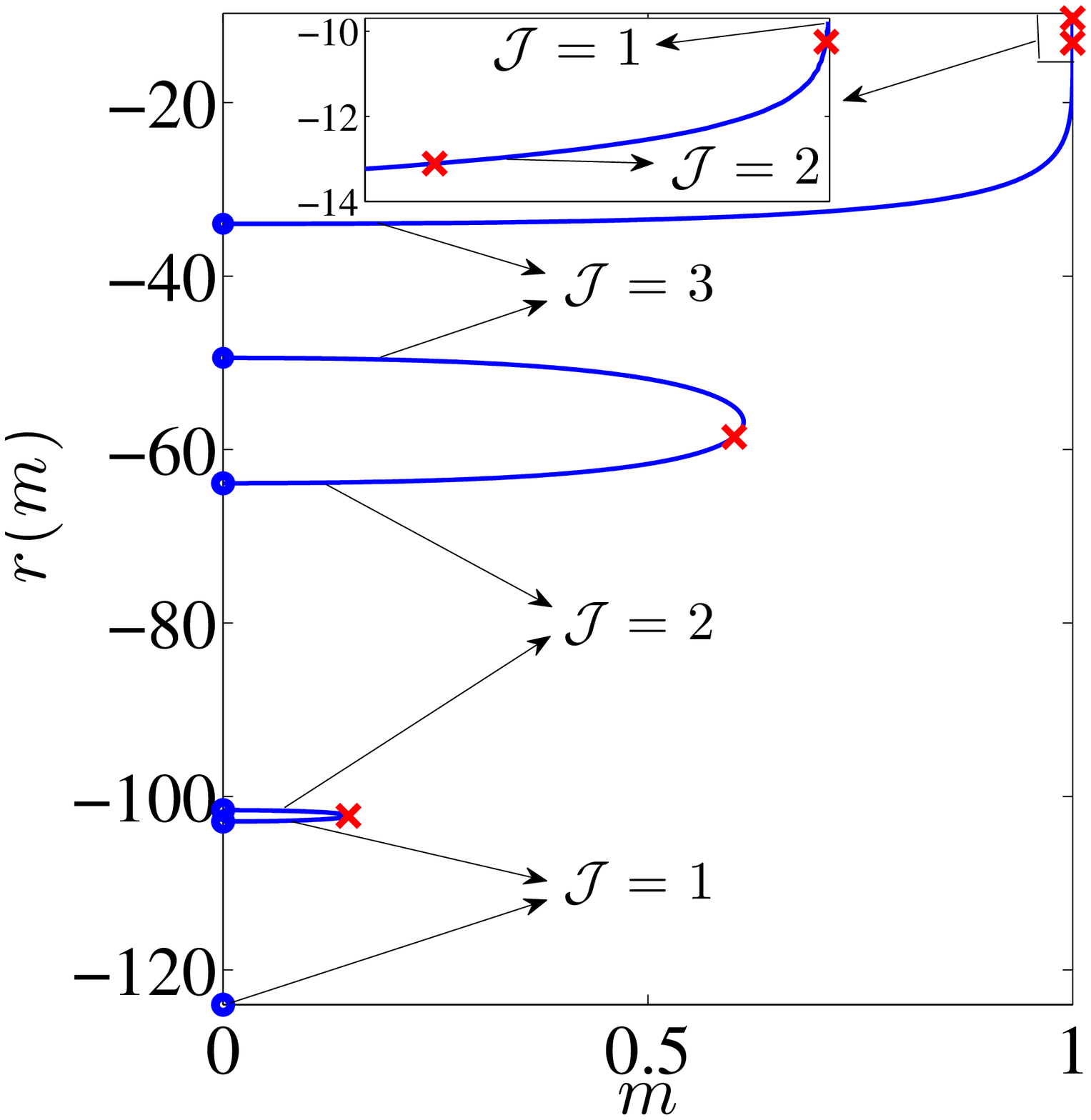}
 \caption
{The ratio $r(m)$ plotted as a function of $m$ for $x_B=0.2$ (left), 
0.04 (middle) and 0.008 (right). The dots on the $r$ axis 
mark the 
values of $r_0(\mathcal J)$ and the crosses mark the location
where 
the phase winding number $J_B$ has a jump of $\pm 1$. The inset in the 
right figure is an expanded view of the 
$r(m)$ curve in the vicinity of $m=1$. The 
interaction parameter is $\gamma = 23$. }
\label{r_vs_m}
\end{center}
\end{figure*}

We thus find that for any $\mathcal J > 1$, negative solutions of 
Eq.~(\ref{r_0}) (always two) exist if  $\gamma> 2\mathcal J(\mathcal 
J-1)$ and if $x_B$ is 
sufficiently small. As $x_B$ increases further, the two negative
roots approach each other and a double negative 
root eventually appears at some critical value $x_B^{\rm 
crit}(\gamma,\mathcal J)$, 
above which no negative root is possible. This behaviour is
again illustrated in Fig.~\ref{Y_r}. The condition
for the existence of a double negative root is given by
 \begin{equation}
 \Delta = 0,
 \label{discriminant}
 \end{equation}
where $\Delta$ is the discriminant~\cite{Irving} of $Y(\tilde r_0)$.  The 
discriminant $\Delta$ is a polynomial function of $x_B$ and
$\gamma$; since it is rather unwieldy, we will not write it out here. 
It is nonetheless straightforward to solve Eq.~(\ref{discriminant})
numerically. The results obtained for $x_B^{\rm crit}(\gamma,
\mathcal J)$ as a function of $\gamma$ for $\mathcal J=2$, 3, 4
and 5 are shown in Fig.~\ref{x_B_crit}. Each of these curves
begins at $\gamma = 2\mathcal J(\mathcal J-1)$ and below this
value, a
double negative root for that particular value of $\mathcal J$
cannot exist for {\it any} value of $x_B$. 
It is convenient to extend the definition of
$x_B^{\rm crit}(\gamma,\mathcal J)$ by defining $x_B^{\rm
crit}(\gamma,\mathcal J)= 0$ for $\gamma < 2\mathcal J(\mathcal
J-1)$. In this way, we can say quite generally that negative
roots do {\it not} exist if $x_B > x_B^{\rm
crit}(\gamma,\mathcal J)$.
Since two negative solutions always exist for
$\mathcal J = 1$, we can simply define
$x_B^{\rm crit}(\gamma,1)=1/2$ for all $\gamma$, where
$x_B = 1/2$ is the maximum value of the minority concentration.
Our calculations show that, for a
fixed $\gamma$, $x_B^{\rm 
crit}(\gamma,\mathcal J)$ diminishes rapidly with increasing
$\mathcal J$. Furthermore, $x_B^{\rm crit}(\gamma,\mathcal J)$
increases monotonically as a function of $\gamma$ and
approaches an 
asymptotic value as $\gamma\rightarrow \infty$. An
analysis of Eq.~(\ref{discriminant}) in
the $\gamma\rightarrow \infty$ limit yields the result
 \begin{equation}
 \lim_{\gamma\rightarrow \infty}x_B^{\rm crit}(\gamma,\mathcal 
J)=\frac{1}{2}-\frac{1}{2}\sqrt{1-\mathcal J^{-2}}.
 \end{equation}

Fig.~\ref{x_B_crit} is extremely useful in determining
what the possible negative roots of Eq.~(\ref{r_0}) are. We
imagine drawing a vertical line in Fig.~\ref{x_B_crit} at some
particular $\gamma$ value. This line will intersect the
$x_B^{\rm crit}(\gamma, \mathcal J)$ curves for $\mathcal J \le
\mathcal J_{\rm max}$. For a given choice of $x_B$, negative
roots occur if $x_B < x_B^{\rm crit}(\gamma,\mathcal J)$, with
$\mathcal J \le \mathcal J_{\rm max}$. This condition will be
satisfied for some maximum winding number $q \le \mathcal J_{\rm
max}$. In other words, negative $r_0(\mathcal J)$ roots will appear
for $\mathcal J  = 1,...,q$. We denote these pairs of roots by
$r_0^<(\mathcal J)$ and $r_0^>(\mathcal J)$, with
$r_0^<(\mathcal J)< r_0^>(\mathcal J)$. In addition, we find
that $r_0^>(\mathcal J)< r_0^<(\mathcal J+1)$, so that the roots
form an increasing sequence which can be plotted along the $m=0$
axis in the $r(m)$-$m$ plane. An example of this is
shown in Fig.~\ref{r_vs_m}. The roots in this sequence are 
bounded from below by $r_0^<(1) = -x_A/x_B$.

The above information can now be used to numerically determine
the negative $r(m)$ solutions of Eq.~(\ref{phasebc'}) for $m\ne 0$.
We find that there exists a unique value of $m$ for each $r$
value in the range $r_0^>(\mathcal J)\le r \le r_0^<(\mathcal
J+1)$ for $\mathcal J = 1,\cdots,q-1$ and $r_0^>(q)\le r \le r_{0,\rm max}$, where 
$r_{0,\rm max}$ is an upper bound whose value depends on $\gamma$ and $x_B$;
 no solution exists for negative $r$ outside one of these ranges.
When plotted in the $r(m)$-$m$ plane, the solutions quite
generally either form a {\it lobe} which starts and ends on the $m=0$
axis or a curve which extends from $m=0$ to $m=m_{\rm max}$. The
number of lobes is given by $q-1$.
This behaviour is illustrated in Fig.~\ref{r_vs_m} for the
example of $\gamma = 23$, in which case ${\mathcal J}_{\rm max}
=3$.

In the left panel of Fig.~\ref{r_vs_m}
($x_B = 0.2$), $q=1$ and no lobe appears in the solution 
of $r(m)$; the phase winding number difference is fixed at 
$\mathcal J = 1$. The isolated point at $r = -x_A/x_B$ on the 
$m=0$ axis corresponds to the $l\to x_B$ limit of the class (i) 
solution which, as we have seen, is $\psi_A = 1/\sqrt{2\pi}$ and
$\psi_B = e^{i(\theta-\theta_0)}/\sqrt{2\pi}$. The continuous
curve is a different soliton branch which at $m=0$
reduces to exactly the same plane wave solution with $l=x_B$. As $m$
increases, $l$ increases continuously from $x_B$ to 1/2. Since
$r(m) > r_0^<(1)$, we have $\gamma_A > 0$ and 
$\gamma_B <0$, which implies that the $A$ component forms a grey 
soliton while the $B$ component forms a bright soliton.

More generally for $q>1$, as illustrated in the middle and right 
panels of Fig.~\ref{r_vs_m}, the two $r_0(\mathcal J)$ negative roots
correspond to the plane wave state $\psi_A = 1/\sqrt{2\pi}$ and
$\psi_B = e^{i\mathcal J(\theta-\theta_0)}/\sqrt{2\pi}$ with
$\mathcal J = 1,...,q$. Each branch of the $r$ {\it vs} $m$
curves, labelled from the bottom
to the top by the index $k=1,...,q$, represents a distinct
soliton state.  On the $k$-th lobe ($k\leq q-1$), the angular
momentum increases from $kx_B$ at the lower end to 
$(k+1)x_B$ at the upper end. On the final $k=q$ branch, the angular 
momentum continues to increase from $l=qx_B$ 
to $l = 1/2$. We emphasize that the disconnected structure of the 
$r(m)$ branches does not imply any discontinuous behaviour of the
condensate wave functions as a function of $l$. In fact, since
$r_0^<(\mathcal J)$ and
$r_0^{>}(\mathcal J)$ correspond to the same plane wave 
functions, $\psi_s=e^{iJ_s(\theta-\theta_0)}$, the $m=0$ wave
functions
at the top of one branch are the same as those at the bottom of
the next. In other words, the various soliton 
branches are always stitched together by plane wave 
functions at angular momenta which are multiples of $x_B$. 

\begin{figure*}[!htbp]
\begin{center} 
 \includegraphics[angle=0, width=2\columnwidth]{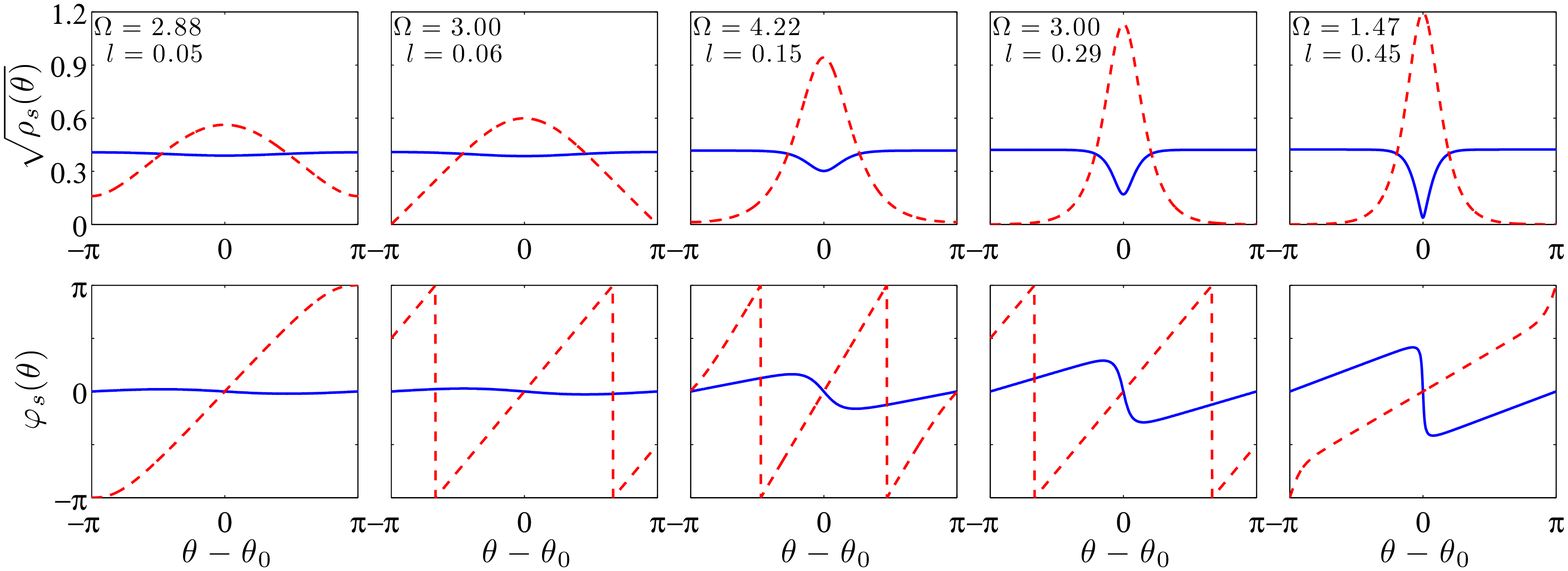}
 \caption
{Soliton amplitude and phase at various angular momenta for $x_B = 0.04$ and 
$\gamma =23$. The solid (blue) and dashed (red) lines represent the 
$A$ and $B$ components, respectively.}
\label{wf}
\end{center}
\end{figure*}

We also find that the phase winding number $J_B$ 
exhibits a rather interesting behaviour as a function of $l$. 
Since the plane wave state of the $B$ component at the
end-points of the $k$-th lobe changes
from $\psi_B=e^{ik (\theta-\theta_0)}/\sqrt{2\pi}$ 
to $\psi_B=e^{i(k+1) (\theta-\theta_0)}/\sqrt{2\pi}$, the phase winding
number $J_B$ must increase by $1$ along this path. This means
that $J_B$ must jump from $k$ to $k+1$ somewhere along the
path. This behaviour is confirmed by our numerical calculations
which show that the jump in $J_B$ occurs at the point along the
path where the $B$-component density develops a node at $\theta
= \theta_0+\pi$, that is, opposite to where its density is a
maximum. These points along the path are indicated by the red
crosses in Fig.~\ref{r_vs_m}.
Finally, on the
$k=q$ branch, $J_B$ decreases from $q$ to 1 via a sequence of
$\theta = \theta_0+\pi$ nodes.  
Thus, the phase of the $B$ component first winds up for $k =
1,...,q-1$ and then unwinds on the $k=q$ branch
as the angular momentum per particle $l$ varies continuously 
from $x_B$ to $1/2$. These features can be seen clearly from the 
density and 
phase plots of the solitons at various angular momenta, shown in 
Fig.~\ref{wf} for $x_B = 0.04$. The figure shows that $J_A = 0$
over this range of angular momenta while $J_B$ starts at 1,
jumps to 2 at $l = 0.06$ and then back to 1 at $l=0.29$. At $l =
1/2$, both components develop a node and $J_A$ jumps from 0 to 1
while $J_B$ jumps from 1 to 0.
\begin{figure}[!htbp]
\begin{center} 
 \includegraphics[angle=0, width=1\columnwidth]{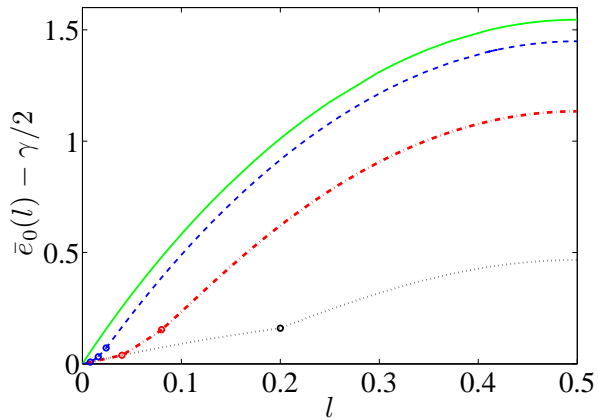}
 \caption
{The yrast spectrum $\bar e_0(l)$ as a function of $l$ plotted for 
various values of $x_B$. The curves from bottom to the top correspond 
to, $x_B = 0.2$ 
(dot, black), $ 0.04$ (dot-dash, red), $ 0.008$ (dash,
blue), and $ 0$ (solid, green). The circles 
mark the locations at which the slope of the curves has a 
discontinuity (see Figure~\ref{e_vs_l_b}). The interaction parameter is 
$\gamma = 23$.  }
\label{e_vs_l}
\end{center}
\end{figure}
\begin{figure}[!htbp]
\begin{center} 
  \includegraphics[angle=0, width=1\columnwidth]{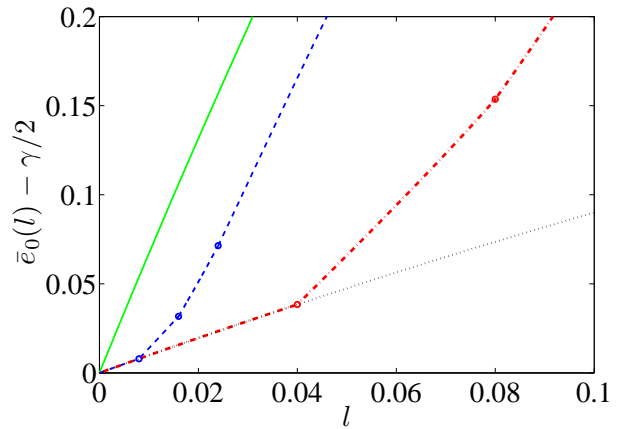}
 \caption
{An expanded view of the small $l$ region of Fig.~\ref{e_vs_l}. }
\label{e_vs_l_b}
\end{center}
\end{figure}

We now use the solution of $r(m)$ for the grey-bright solitons
to evaluate the mean-field yrast spectrum from Eq.~(\ref{E_total}). 
In Fig.~\ref{e_vs_l} we plot $\bar e_0(l)-\gamma/2$ for $\gamma =23$
and various concentrations, $x_B$. We see that the yrast spectrum
is a continuous function of $l$ but consists of several segments
corresponding to the different soliton states that occur. The
first segment for $0\le l \le x_B$ is the energy of the class (i)
solution, which is given by $\bar e_0(l) = 
l - l^2+\gamma/2$ (see Eq.~(\ref{E0ni})). The remaining
segments correspond to the distinct $r(m)$ branches in 
Fig.~\ref{r_vs_m}. Generally speaking, there are $q-1$ segments,
each of width $x_B$, associated with the lobes in 
Fig.~\ref{r_vs_m} and a final segment extending from $l=
qx_B$ to $l=1/2$. Although the energy $\bar e_0(l)$ is
a continuous function of $l$, its derivative, which is related to 
the angular velocity $\Omega$ through $d\bar e_0 (l)/dl =\Omega - l$, 
is not. The discontinuities occur at $l = kx_B$ ($k =
1,...,q$), the end-points of the various segments. As a result,
the angular velocity $\Omega$ has different limiting values as
the angular momentum approaches one of these points from either
side. More quantitatively, using
Eq.~(\ref{phasebc}) we find
that the two angular velocity limits at $l=kx_B$ are
\beq
\Omega^{\gtrless}_k = \sqrt{1 + 2\gamma_{A}^{\gtrless}(k)}\,,
\label{omega_pm}
\eeq
where $\gamma_{A}^{\gtrless}(k)=\gamma \left [x_A + x_B
r^{\gtrless}_0(k) \right ]$ and 
$<(>)$ denotes the limit as $l$ approaches $kx_B$ from lesser
(greater) values.
In deriving this result we made use of the fact
that ${\rm sgn}W_A=-1$ in Eq.~(\ref{phasebc}). 
To see this, we observe that
$\calM_A\geq \calM_B >0$ for $\gamma_A > 0$ and $\gamma_B < 0$
(see
Appendix). Using this property in Eq.~(\ref{phasebc'}) and
bearing in mind
that $\mathcal J = J_B-J_A >0$, we find $ {\rm sgn}W_A = -1$.
From
Eq.~(\ref{omega_pm}) we also find the following inequalities
\beq
1 \le \Omega_{k-1}^< \le \Omega_{k-1}^> \le \Omega_k^< \le
\Omega_k^>
%\Omega^>_{i}>\Omega^<_{i}>\Omega^>_{i-1}>\Omega^<_{i-1}\geq 1,
\label{omega_iq}
\eeq
which will be useful in the discussion of persistent currents in
the next section.

\section{Persistent current at higher angular momentum}
In this section, we investigate within mean-field theory the 
possibility of persistent currents of the two-component system at higher 
angular 
momenta. For a single component repulsive gas (the $x_B=0$ limit of 
the two-component system), persistent currents are
sustainable at any integer angular momentum $l=\nu$ provided the 
interaction strength $\gamma$ exceeds a critical value given 
by~\cite{Anoshkin}
\beq
\gamma = \frac{4\nu^2 -1 }{2}.
\label{gs_cr}
\eeq
Although the addition of a second component has the tendency of 
destabilizing persistent currents at higher angular
momenta, it does not completely remove the possibility, contary 
to the suggestion made in Ref.~\cite{Smyrnakis1}. 
In support of this we presented in an earlier paper~\cite{Anoshkin} a 
general argument based on the idea that the energy 
spectrum of the two-component system should continuously reduce to the 
single component result as the concentration of the minority component 
$x_B$ gradually goes to zero. The purpose of this section is to use the 
exact soliton solutions for the yrast spectrum to provide a more 
definitive, and quantitative, argument for why persistent
currents at higher angular momenta are indeed possible.

The possibility of persistent currents is predicated on the existence of 
local minima in the yrast spectrum. For a single component gas, local 
minima (if they exist) always occur at integer 
angular momenta per particle. This is no longer the case for a 
two-component system. In fact, as a result of the non-interacting-like 
spectrum for $0\leq l\leq x_B$, local minima of the yrast spectrum {\it 
cannot}  appear at integer values of angular momentum $l$. However, as 
a key result of this section, we will demonstrate that local minimum 
can form at angular momenta $l = \nu x_A=\nu (1-x_B)$ for any integer 
$\nu$, provided that $\gamma$ is greater than the critical strength
\beq
\gamma_{cr,\nu} = \frac{4\nu^2 - 1}{2(1 - 4x_B\nu^2)}.
\label{gamma_cr}
\eeq  
This proves $\it unequivocally$ that persistent currents are possible at 
arbitrarily high angular momentum.  We observe that in the
$x_B\rightarrow 0$ limit, $\gamma_{cr,\nu}$ in fact reduces to the 
single-component critical criterion given in
Eq.~(\ref{gs_cr}).  A corollary of Eq.~(\ref{gamma_cr}) is that
a local minimum at $l = \nu x_A$ cannot arise for any $\gamma$ if 
\beq
x_B > \frac{1}{4\nu^2}.
\label{xbcritnu}
\eeq

To derive the result in Eq.~(\ref{gamma_cr}), we focus on the angular 
velocity of the solitons as a function of $l$ which is equal to the 
slope of the mean-field yrast spectrum $\bar E_0(l)$. We will be 
particularly interested in parameter regimes
where the angular velocity develops discontinuities at $l = k x_B$. 
From the periodicity of the yrast spectrum,
the angular velocity will also have discontinuities at
$l = \nu - kx_B$. Our objective is to investigate the
possibility of the formation of a local minimum at one of these angular 
momenta. The condition for this to occur is that the right limit of the 
angular velocity is positive and the left limit is negative.  

In the argument to be presented, we require the index $k$ to span a range that includes the integer $\nu$. To ensure this,
we assume that $\gamma > 2\nu(\nu-1)$ and that $x_B < x_B^{\rm
crit}(\gamma,\nu)$ so that the number of discontinuities is $q
\ge \nu$. Due to 
the inversion symmetry and periodicity of the yrast spectrum, the 
angular velocity at $\nu - l$ is related to that at $l$ through
\beq
\Omega(\nu - l) =2\nu-\Omega(l). 
\label{Omegap}
\eeq
This implies that the left limit of the angular velocity at $l =
\nu - kx_B$ is 
\beq
\Omega^<(\nu - kx_B) =2\nu-\Omega^>_k, 
\label{omega<}
\eeq
while the right limit is
\beq
\Omega^>(\nu - kx_B) =2\nu-\Omega^<_k, 
\label{omega>}
\eeq
where $\Omega_k^\gtrless$ is given in Eq.~(\ref{omega_pm}).
In view of Eq.~(\ref{omega_iq}), we see that $\Omega^<(\nu-(k+1)x_B) 
< \Omega^<(\nu-kx_B)$, that is, the left limit decreases with
increasing $k$. We will prove that $\Omega^<(\nu-kx_B)$ is
positive for $k<\nu$ and that $\Omega^<(\nu-\nu x_B)$ is the
first term in the sequence which can become negative with
increasing $\gamma$. This occurs when $\Omega^>_\nu > 2\nu$. 
We thus focus
on $\Omega^\gtrless_\nu$ in the following.%=2\nu -\Omega^\lessgrt_\nu$.

%we have $\bar e_0(\nu-l)=-\bar e_0(l)$, where $0\leq l\leq 1/2$. 
%In view of Eq.~(\ref{omega}) and the fact that 
%$\bar E_0(l)=\bar e_0(l) + l^2$, we find that the angular 
%velocity at $l = \nu x_A$ is related to that at $\nu x_B$ through
%\beq\Omega(\nu x_A) =2\nu-\Omega(\nu x_B). \eeq
The quantities $\Omega^{\gtrless}_\nu$ are given by 
Eq.~(\ref{omega_pm}), however, to use this result we need to 
solve Eq.~(\ref{r_0}) 
for $r_0$. An alternative, and more convenient approach is to
obtain an equation that determines $\Omega^{\gtrless}_\nu$
directly. By setting $m = 
0$ and $\mathcal J = \nu$ in Eq.~(\ref{phasebc'}) and using 
Eq.~(\ref{phasebc}) to eliminate $r_0$ in favour of $\Omega$, we find 
that the desired equation is
\begin{equation}
\left (\Omega^2 -1 - 2\gamma \right )\left [(\Omega - 2 \nu)^2 -1 
\right ] + 8 \gamma x_B \nu(\nu-\Omega) =0.
\label{eqomega}
\end{equation}
We now make a few key observations regarding the solutions to 
Eq.~(\ref{eqomega}). First, for $\gamma$ and $x_B$ within the 
aforementioned range, Eq.~(\ref{eqomega}) must yield two positive 
roots given by Eq.~(\ref{omega_pm}).  In addition, the quartic
in Eq.~(\ref{eqomega}) is negative at $\Omega=0$, which along 
with the first observation guarantees that Eq.~(\ref{eqomega}) has four 
{\it real} solutions, one negative and three positive. Finally, 
one can show that the two {\it smallest} positive roots are associated
with the two negative $r_0$ roots of Eq.~(\ref{r_0}) and thus
can be identified with $\Omega^{\gtrless}_\nu$.

\begin{figure}[!htbp]
\begin{center} 
 \includegraphics[angle=0, width=1\columnwidth]{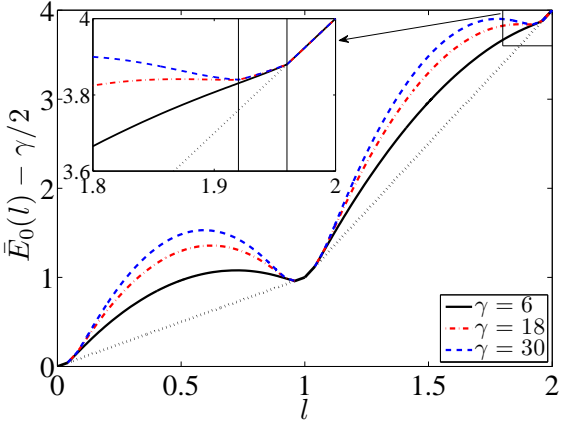}
 \caption
{Mean-field yrast spectrum for $\gamma = 6$, 18 and 30 at $x_B=0.04$. The 
inset is an expanded view of the behaviour of the energy curves
in the vicinity of $l=2$. The linear lines from $l=0$ to $1$ and from $1$ to $2$ 
correspond to the non-interacting yrast spectrum.}
\label{yrast_2}
\end{center}
\end{figure}

\begin{figure}[!htbp]
\begin{center} 
 \includegraphics[angle=0, width=1\columnwidth]{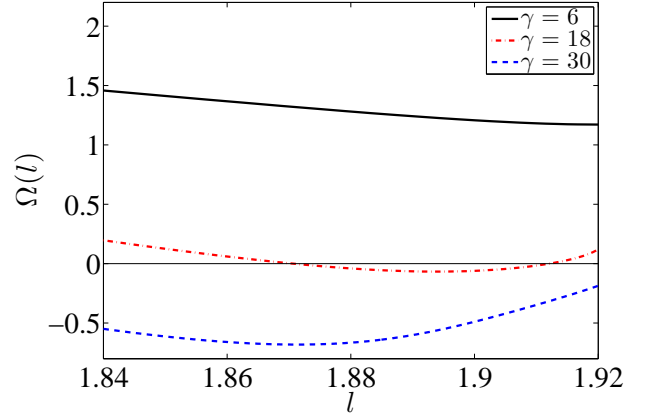}
 \caption
{The angular velocity $\Omega(l)$ plotted as a function of $l$ for 
$l<2-2x_B$.}
\label{Omega}
\end{center}
\end{figure}
To see where $\Omega^{\gtrless}_\nu$ are located, we first consider the 
$x_B = 0$ limit, in which case the solutions to Eq.~(\ref{eqomega}) are 
$\Omega = 2\nu \pm 1$ and $\Omega=\pm \sqrt{1 + 2\gamma}$. Two of these 
four solutions correspond to $\Omega^{\gtrless}_\nu$ in the limit that 
$x_B \rightarrow 0$. 
Since $2\nu-1$ is the 
smallest positive solution for $\gamma > 2\nu (\nu-1)$,
we see that $\lim_{x_B\rightarrow 0}\Omega_\nu^< = 2\nu-1$.
As for $\lim_{x_B\rightarrow 0}\Omega_\nu^>$, it takes the value 
${\rm min}\{2\nu+1,\sqrt{1+2\gamma}\}$. Thus, if
$2\nu (\nu-1)<\gamma \leq 2\nu (\nu+1)$,  $\lim_{x_B\rightarrow
0}\Omega_\nu^> = \sqrt{1+2\gamma}$. This implies that
$\lim_{x_B\rightarrow 0}\Omega_\nu^> > 2\nu$ if $\gamma > 
(4\nu^2 - 1)/2$, which is the critical interaction
strength for $x_B =0$ given in Eq.~(\ref{gs_cr}). For $x_B\ne
0$, the critical interaction strength can be obtained simply by
inserting $\Omega = 2\nu$ into Eq.~(\ref{eqomega}) and solving
for $\gamma$. This leads to the result given in Eq.~(\ref{gamma_cr}).
For $\gamma$ slightly larger than $\gamma_{{\rm cr},\nu}$, we
have $\Omega^<(\nu-\nu x_B) < 0$ and, since $\Omega_\nu^< <
\Omega_\nu^>$, $\Omega^>(\nu-\nu x_B) > 0$. In other words, we
have established the possibility of a local minimum at $l=\nu
x_B$.

Finally, we demonstrate that $\Omega^<(\nu-k x_B) > 0$ for
$k<\nu$ and for {\it any} $\gamma$ larger than $2\nu(\nu-1)$. Using
the argument given above, we have 
$\lim_{x_B\rightarrow 0}\Omega_k^> =
{\rm min}\{2k+1,\sqrt{1+2\gamma}\}$. If $\gamma > 2\nu(\nu-1)$,
$\lim_{x_B\rightarrow 0}\Omega_k^> = 2k+1$. Furthermore,
it is not difficult to see from 
Eq.~(\ref{eqomega}) that $\Omega_k^> <\lim_{x_B\rightarrow 
0}\Omega_k^> = 2k+1$. Inserting this result into
Eq.~(\ref{omega>}), we see that $\Omega^<(\nu -k x_B) > 2\nu - 
(2k+1) > 0$ if $k<\nu$. This proves that a local minimum cannot
occur at $l = \nu- kx_B$ for $k<\nu$ for {\it any} value of
$\gamma$. This result was established earlier for the case $\nu
= 2$ and $k=1$~\cite{Smyrnakis1, Anoshkin}.

To close this section, we emphasize that although a 
local minimum at $l = \nu x_A$ cannot occur if
$\gamma < \gamma_{cr,\nu}$,
it  does {\it not} necessarily preclude the possibility of a
local minimum at locations away from $l = \nu x_A$. To demonstrate 
this, we consider $x_B=0.04$ as an example. According to 
Eq.~(\ref{xbcritnu}), local minima are possible at $l=\nu - \nu x_B$ 
for $\nu=1$ and $2$.  Let us focus on $\nu = 2$. From 
Eq.~(\ref{gamma_cr}) we see that the critical $\gamma$ value below 
which no local minimum appears at $l=2-2x_B$ is
$\gamma_{cr,2}=125/6 \simeq 20.8$. 
In Fig.~\ref{yrast_2} we have obtained the mean-field yrast spectrum 
for a series of $\gamma$ values below and above this critical value.
The qualitative behaviour proposed in our earlier work~\cite{Anoshkin} 
is completely consistent with the quantitative results found here. The
corresponding slope of the spectrum for $l \le 2-2x_B$ is shown in 
Fig.~\ref{Omega}. For $\gamma =30 >\gamma_{cr,2}$, it is clear that the 
slope of the spectrum approaching $l=2-2x_B$ from the left side is 
negative and there is a local minimum at this angular momentum value. 
For $\gamma =18 <\gamma_{cr,2}$, however, the slope at
$l=2-2x_B$ is {\it positive}, 
implying that the yrast spectrum does not have a local minimum
at this angular momentum for this $\gamma$ 
value. But as we can see from Fig.~\ref{Omega}, the slope decreases as 
$l$ moves away from $l=2-2x_B$ and eventually passes through
zero, indicating that a local minimum of the yrast 
spectrum occurs at some angular momentum $l < 2-2x_B$.

\section{Conclusions}
In this paper, we have presented a comprehensive analysis
of the mean-field yrast spectrum of a two-component gas in the
ring geometry, using the analytic soliton solutions found in
Refs.~\cite{Porubov,Smyrnakis3,Smyrnakis2}. We find that the yrast
 spectrum reveals a complicated sequence of soliton states  
 as a function of the interaction strength $\gamma$ 
 and the population imbalance, specified by
the minority concentration $x_B$. 
A denumerably infinite set of  
$x_B^{\rm crit.}(\gamma,\mathcal J)$ curves divides 
the $\gamma-x_B$ plane into regions, 
within each of which the yrast spectrum
has a distinct structure. Our findings also demonstrate
that the mean-field yrast spectrum of the two-component gas 
reduces to that of the single-component system in a highly 
non-trivial way as $x_B$ tends to zero.
Finally, the analysis of the yrast spectrum allows us to provide a
definitive answer to the question of whether persistent currents at 
higher angular momenta can occur. In particular, we have derived
 an analytic formula for the critical interaction strength, 
above which persistent currents are found to exist at $l=\nu(1-x_B)$
 for arbitrary integer $\nu$. It would clearly be of interest to see if these
 theoretical predications can be verified experimentally. \appendix
\section{Soliton solutions for the single-component gas}
Knowledge of the single-component soliton solutions (both attractive and repulsive) is necessary to understand the two-component case. In this appendix, 
we provide some useful information on these solutions, particularly on the evaluation of the quantities $\mathcal M_s$ and $W_s$ in Eqs.~(\ref{phase})-(\ref
{phasebc}). Much of the material provided here can also be found in Refs.~\cite{Carr1,Carr2,Kanamoto1,Kanamoto3}. 

The densities and phases of the single component solitons are given by Eqs.~(\ref{rhoa}) and (\ref{phase}), which contain an unspecified elliptic parameter 
$m$. We will see that $m$ cannot assume all the values from $0$ to $1$ for either attractive or repulsive interactions. The allowable range of $m$ is 
determined by the requirement that the quantity $W_s^2$ is non-negative. For clarity we discuss seperately the case of attractive and repulsive interactions.
\subsection{Attractive interaction}
\begin{figure}[!htbp]
\begin{center} 
 \includegraphics[angle=0, width=0.43\columnwidth]{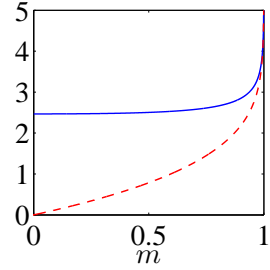}
 \caption
 {$KE$ (solid line) and $KE-(1-m)K^2$ (dashed line) as a function of $m$.}
\label{KE1}
\end{center}
\end{figure}

For $\gamma_s<0$, we see from its definition that $f_s < 0$ since $KE-K^2\leq 0$. In view of Eq.~(\ref{Wsqr}) and the fact that $h_s \leq g_s$, we 
must have $h_s\leq 0$ and $g_s \geq 0$ in order that $W_s^2 \geq 0$. This leads to the inequality
\begin{equation}
KE-(1-m)K^2  \leq  \frac{\pi^2}{2j^2}|\gamma_s|\leq KE.
\label{ineq2}
\end{equation}
Since the bounds of this inequality are monotonically increasing functions of $m$ (see Fig.~\ref{KE1}), one finds
\begin{equation}
m_{\rm min}\leq m\leq m_{\rm max},
\label{m_range}
\end{equation}
where $m_{\rm max}$ is the solution of equation $h_s = 0$, while $m_{\rm min}=0$ if $|\gamma_s|<2j^2 K(0)E(0)/\pi^2=j^2/2$, and is otherwise the solution 
of the equation  $g_s = 0$.

To evaluate the quantity $\mathcal M_s$ for the allowed values of $m$ given by (\ref{m_range}), we need the complete elliptic integral of the third kind $\Pi\left
(n_s;K|m\right )\equiv \Pi\left(n_s\backslash\alpha \right )$, where $\alpha=\sin^{-1}\sqrt{m}$. For $\gamma_s <0$, one finds $m<n_s<1
$ and the elliptic integral can be evaluated as~\cite{Abramowitz}
\begin{equation}
\Pi(n_s\backslash \alpha) = K +\frac{\pi}{2}\delta_2[1-\Lambda_0(\epsilon_s\backslash \alpha)],
\label{comellip3}
\end{equation}
where 
$\delta_2 = jK\sqrt{{2f_s}/{(g_sh_s)}}$, $\epsilon_s=\sin^{-1}\sqrt{{h_s}/{[f_s(1-m)]}}$ and $\Lambda_0$ is Heuman's Lambda function, defined as
\begin{equation}
\Lambda_0(\epsilon_s\backslash \alpha)=\frac{2}{\pi}\left \{K E(\epsilon_s\backslash \pi/2-\alpha) -[K-E]F(\epsilon_s\backslash \pi/2-\alpha)\right \}.
\label{lamb0}
\end{equation}
Here $F(\phi\backslash\alpha)$ and $E(\phi\backslash\alpha)$ are incomplete elliptic integrals of the first and second kind.
Using Eq.~(\ref{comellip3}), one finds~\cite{Kanamoto1}
\begin{align}
\calM_s(m)=\frac{1}{2\pi}\left \{\sqrt{{2g_sh_s}/{f_s}}+{j}\pi\left [1-\Lambda_0(\epsilon_s\backslash \alpha)\right ]\right \}.
\label{findkn}
\end{align}

\begin{figure}[!htbp]
\begin{center} 
 \includegraphics[angle=0, width=1\columnwidth]{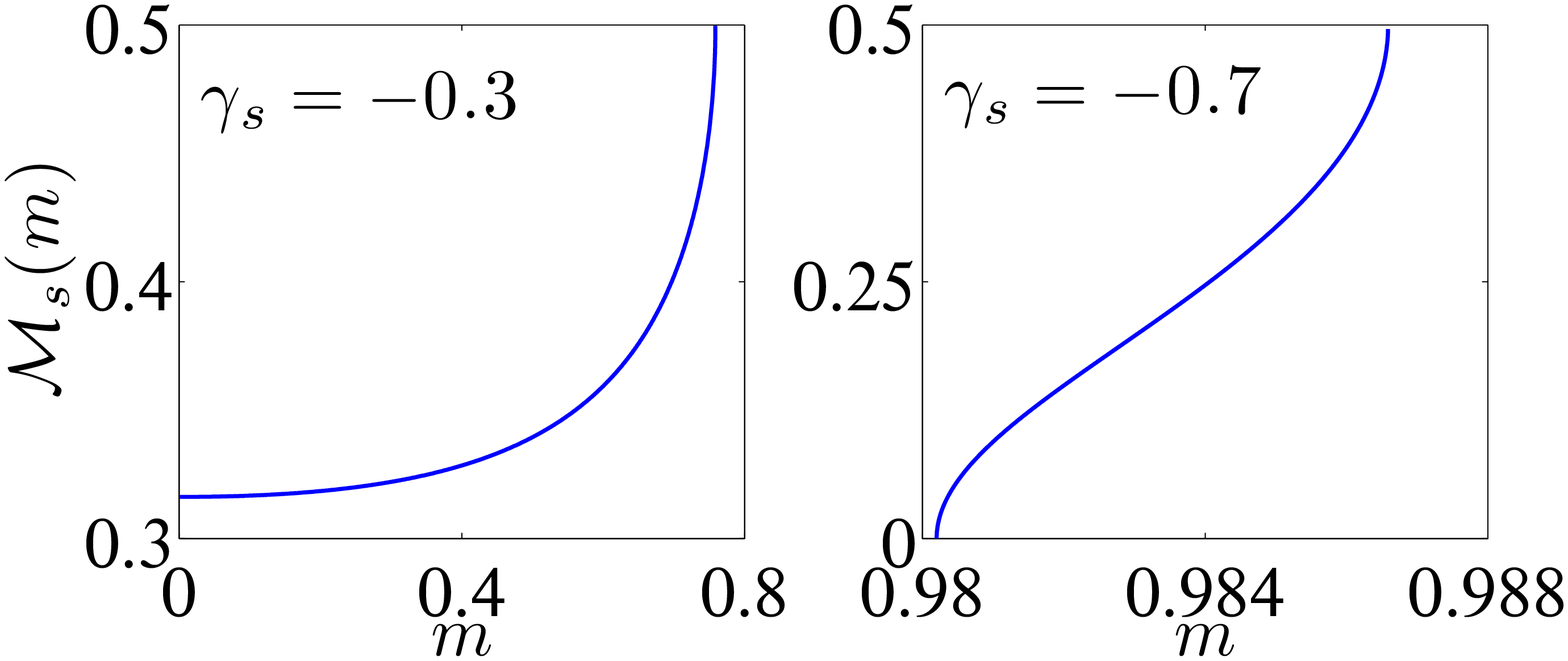}
 \caption
 {$\calM_s (m)$ ($j=1$) as a function of $m$ for $\gamma_s=-0.3$ and $\gamma_s=-0.7$.}
\label{M_att}
\end{center}
\end{figure}

To be specific, we now consider the case of $j=1$, namely a single soliton. In this case $\calM_s (m)$ is a monotonically increasing function of $m$. At 
$m=m_{\rm max}$, we have $h_s=0$ which leads to $\e_s=0$ and $\Lambda_0(\e_s\backslash\alpha)=0$.  Thus we find from Eq.~(\ref{findkn})  that  $
\calM_s (m_{\rm max})=1/2$. Using Eq.~(\ref{Wsqr}) we also find that $W_s(m_{\rm max})=0$.
The lower bound of $\calM_s(m)$ depends on the value of $\gamma_s$. For $|\gamma_s|\leq 1/2$, $m_{\rm min}=0$. At this value we have $g_s=
\pi^2\gamma_s+\pi^2/2$ and $f_s=h_s=\pi^2\gamma_s$. Thus we find $\e_s=\pi/2$ and $\Lambda_0=1$. Equation (\ref{findkn}) then gives  $\calM_s(0)=
\sqrt{2\gamma_s+1}/2$. From Eq.~(\ref{Wsqr}) we find that $|W_s(0)|=\sqrt{2\gamma_s+1}/2\pi$. Now, for $|\gamma_s|>1/2$, $m_{\rm min}$ is the solution 
to $g_s=0$. At this value we have 
$\sqrt{{h_s}/{[f_s(1-m)]}}=1$ and thus $\e_s=\pi/2$. Using the Legendre's relation 
\begin{equation}
EF\left (\frac{\pi}{2}\backslash\frac{\pi}{2}-\alpha\right )+E\left (\frac{\pi}{2}\backslash\frac{\pi}{2}-\alpha\right )K-KF\left(\frac{\pi}{2}\backslash\frac{\pi}{2}-
\alpha\right)=\pi/2,
\end{equation}
we find $\Lambda_0(\e\backslash\alpha)=1$. Thus we obtain $\calM_s (m_{\rm min})=0$ and $W_s(m_{\rm min})=0 $ for $|\gamma_s|>1/2$. As a concrete
example, we show $\calM_s$ calculated for $\gamma_s=-0.3$ and $\gamma_s=-0.7$ in Fig.~\ref{M_att}.

\begin{figure}[!htbp]
\begin{center} 
 \includegraphics[angle=0, width=1\columnwidth]{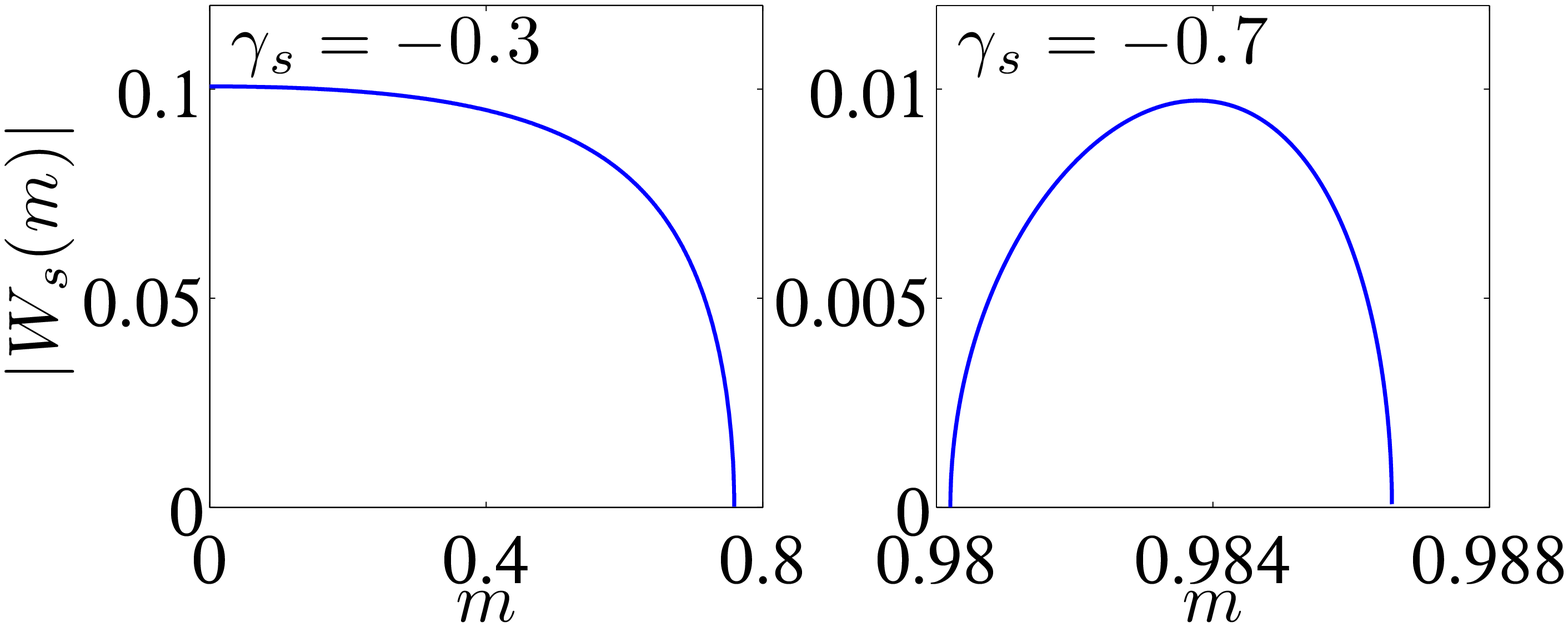}
 \caption
 {$|W_s (m)|$ ($j=1$) as a function of $m$ for $\gamma_s=-0.3$ and $\gamma_s=-0.7$.}
\label{W_att}
\end{center}
\end{figure}

Once $\calM_s$ and $W_s$ have been determined, various physical quantities such as energy, angular momentum and angular velocity can be calculated using the formulae given in Sec. II. For example, according to Eq.~(\ref{phasebc}) and the fact that $0\leq \calM_s\leq 1/2$ 
for $\gamma_s<0$, the range of angular velocity values is given by $2J_s-1\leq \Omega \leq 2J_s +1$. This implies the following important relation for the attractive 
gas~\cite{Kanamoto3}
\begin{equation}
J_s=\left [\frac{\Omega}{2}+\frac{1}{2} \right ],
\end{equation}
where $[x]$ denotes the largest integer less than or equal to $x$. 

\subsection{Repulsive interactions}
For $\gamma_s >0 $, and using the fact that $KE-(1-m)K^2\geq 0$, we find from their definitions that $g_s\geq h_s >0$. Because $h_s \geq f_s$, we must impose the condition $f _s
\geq 0$ in order for $W_s^2 \geq 0$, namely $\pi^2\gamma_s+2j^2KE-2j^2K^2 \geq 0$. In view of the fact that $K^2 - KE$ is a monotonically increasing 
function of $m$, we obtain the allowable range of $m$ as $0\leq m\leq m_{\rm max}$, where $m_{\rm max}$ is the solution to the equation $f_s=0$. 

\begin{figure}[!htbp]
\begin{center} 
 \includegraphics[angle=0, width=1\columnwidth]{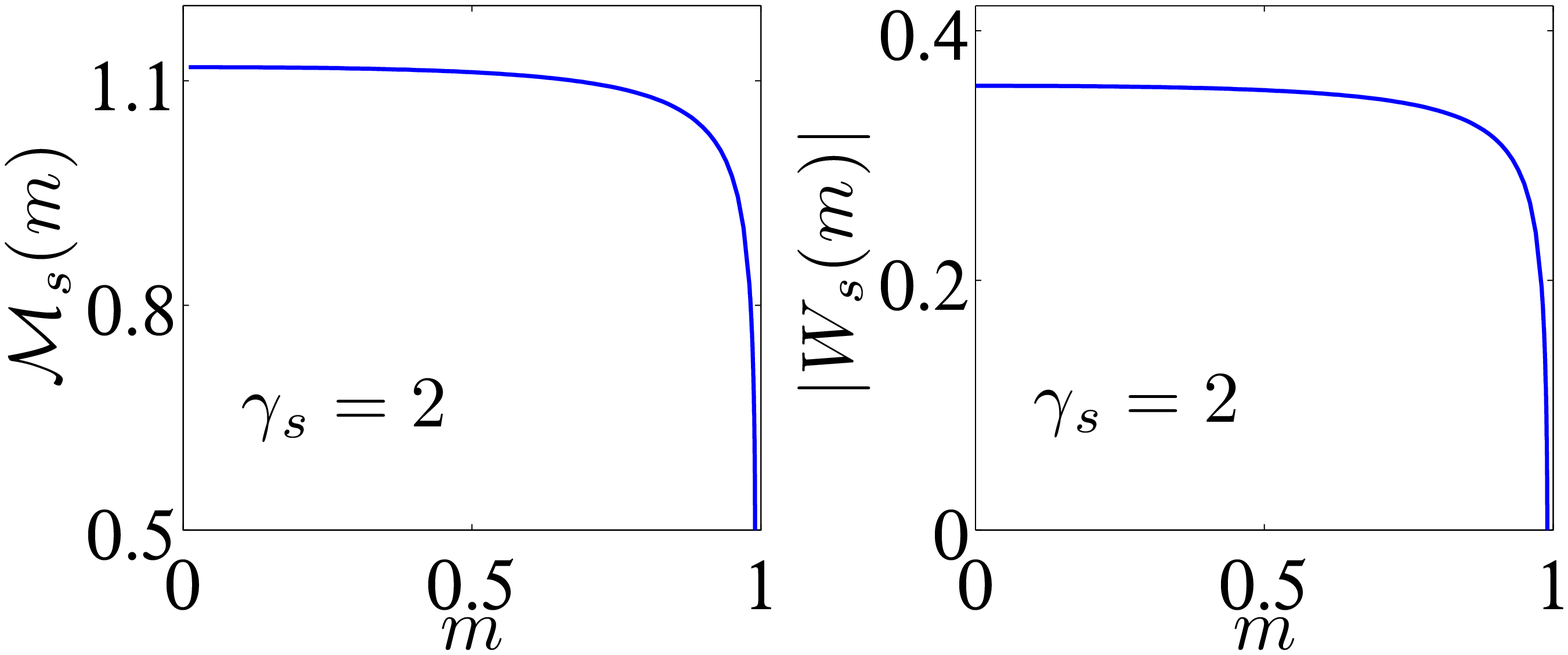}
 \caption
 {$\calM_s (m)$ (left) and $W_s(m)$ (right) as a function $m$ for $j=1$ and $\gamma_s = 2$. }
\label{M_rep_and_l}
\end{center}
\end{figure}

Now, in the case of $\gamma_s > 0$, we have $n_s< 0$. The evaluation of $\Pi(n_s\backslash \alpha)$ for $n_s < 0$ can be reduced to 
the case of $m< n_s'<1$~\cite{Abramowitz}, where $n_s'\equiv(m - n_s)(1-n_s)^{-1}=m{g_s}/{h_s} $. One finds 
\begin{align}
\Pi(n_s\backslash \alpha)
=K\frac{f_s}{g_s}+2j^2(1-m)K^2\frac{f_s}{g_sh_s}\Pi(n_s'\backslash\alpha),
\label{comellip3p}
\end{align}
where $\Pi(n_s'\backslash\alpha)$ can be evaluated as before and is given by
\begin{equation}
\Pi(n'_s\backslash\alpha)= K +\frac{\pi}{2}\delta'_2[1-\Lambda_0(\e'_s\backslash \alpha)].
\label{Pin'}
\end{equation}
Here $\delta'_2=\sqrt{{g_sh_s}/{2f_s}}/{[j(1-m)K]}$ and $\e'=\sin^{-1}\sqrt{{f_s}/{h_s}}$.
Using Eqs.~(\ref{comellip3p}) and (\ref{Pin'}) we obtain for $\gamma_s > 0$
\begin{equation}
\calM_s (m)=\frac{1}{2\pi}\left \{\sqrt{{2g_sf_s}/{h_s}}+j\pi [1-\Lambda_0(\e'_s\backslash\alpha)]\right \}.
\label{findkp}
\end{equation}

The quantity $\calM_s(m)$ for the repulsive gas is a monotonically decreasing function of $m$. For $j=1$ and at $m=0$, we have $f_s=h_s=\pi^2\gamma_s$ 
and $g_s=\pi^2\gamma_s+\pi^2/2$. Using these values we find $\e'_s=\pi/2$ and $\Lambda_0(\e'_s\backslash\alpha)=1$. This leads to $\calM_{s}
(0)=\frac{1}{2}\sqrt{2\gamma_s +1}$. At $m=m_{\rm max}$, we find $f_s=0$, $\e'_s=0$ and $\Lambda_0(\e'_s\backslash\alpha)=0$. Thus we obtain $
\calM_s(m_{\rm max})={1}/{2}$. Similarly we find $0\leq |W_s(m)|\leq \sqrt{2\gamma_s +1}/2\pi$. An illustration of the behaviour of $\calM_s(m)$ and $|W_s
(m)|$ for $\gamma_s=2$ is shown in Fig.~\ref{M_rep_and_l}.

As in the case of attractive interaction, various physical quantities of interest can be obtained after $\calM_s(m)$ and $|W_s(m)|$ have been determined. Here we 
only point out an oversight made in Ref.~\cite{Kanamoto1} regarding the range of the angular velocity $\Omega$.  We first observe that although the phase winding 
number $J_s$ in Eq.~(\ref{phasebc}) can take any integer value, one can without loss of generality restrict its value to $J_s=0,1$, which can be shown to 
correspond to the solutions with angular momenta in the range $0\leq l\leq 1$. Taking $J_s=0$ and ${\rm sgn W_s}=-1$ (${\rm sgn W_s}=1$ renders $\Omega$ 
and $l_s$ negative ), we find from {Eq.~(\ref{phasebc})} that $1\leq\Omega=2\calM_s (m)\leq \sqrt{2\gamma_s+1}$. Since $-\sqrt{2\gamma_s+1}/2\pi \leq 
W_s \leq 0$, we find $ 0\leq l_s=\pi W_s +{\Omega}/{2}\leq {1}/{2}$ for $J_s=0$, where $l_s=0$ at $m=0$ and $l_s=1/2$ at $m=m_{\rm max}$. A similar 
analysis of $J_s =1$ yields $2-\sqrt{2\gamma_s + 1}\leq \Omega \le 1$ for $1/2<l_s \leq 1$.  We thus obtain $2-\sqrt{2\gamma_s+1}\leq\Omega\leq \sqrt
{2\gamma_s+1}$ for $0\leq l_s \leq 1$. This shows that the range of $\Omega$ corresponding to an integral interval of $l_s$ in fact depends on $
\gamma_s$ and in general cannot be restricted to $0\leq \Omega \leq 2$ as claimed in Ref.~\cite{Kanamoto1}. 

Lastly, we provide a useful discussion of  the soliton wave functions at $l_s=0$ and $l_s=1/2$. For the former we have $m=0$, $\Omega=\sqrt{2\gamma_s+1}$ and $J_s=0$. Since ${\rm dn}(u|m=0)=1$, we find from Eq.~(\ref{rhoa}) that $\rho_s(\theta)=1/2\pi$. Furthermore, at $m=0$ one finds $\Pi\left(n_s;{K}(\theta-\theta_0)/\pi,m\right ) =({\theta-\theta_0})/{2}$. Using this result in Eq.~(\ref{phase}) we obtain $\varphi_s(\theta)-\varphi_s(\theta_0)=l_s(\theta -\theta_0)=0$. Thus the wave function is simply $\psi_s(\theta)=1/\sqrt{{2\pi}}$. This is the expected plane wave solution at $l_s=0$. For $l_s=1/2$, we have $m=m_{\rm max}$, $\Omega=1$ and $J_s=0$. Since in this case $f_s=0$ and $g_s=2K^2$, we find from Eq.~(\ref{eta_1}) that $\eta_s=-1$. The density is thus given by
\beq
\rho_s(\theta)=\frac{K^2}{\pi^3\gamma_s}\left [1-{\rm dn}^2\left(\left. \frac{jK}{\pi}(\theta-\theta_0)\right |m\right)\right ].
\eeq
Since ${\rm dn}(u=0|m)=1$, we see that at $\rho_s(\theta_0)=0$ which implies that this is a dark soliton. 

\section*{Acknowledgment}
This work was supported by a grant from the Natural Sciences and Engineering Research Council of Canada.

\end{document}